\documentclass[prb,aps,amssymb,shownopacs,twocolumn,10pt]{revtex4-1}

\usepackage{amsmath}
\usepackage{amssymb}
\usepackage{amsthm}
\usepackage{amsfonts}
\usepackage{listings}
\usepackage{easybmat}
\usepackage{enumerate}
\usepackage{latexsym}
\usepackage{epstopdf}
\usepackage{psfrag}
\usepackage{color}
\usepackage{bm}
\usepackage[pdftex]{graphicx}
\usepackage{hyperref}
\usepackage{ulem} 
\def\upT{\bigtriangleup}
\def\dwT{\bigtriangledown}
\newcommand{\beq}{\begin{equation}}
\newcommand{\eneq}{\end{equation}}

\newcommand{\ket}[1]{\left|#1\right\rangle}

\def\be{\begin{equation}}
\def\ee{\end{equation}}
\def\ba{\begin{eqnarray}}
\def\ea{\end{eqnarray}}

\def\mbf{\mathbf}

\input{epsf}

\begin{document}

\title{Stability of the spin-$1/2$ kagome ground state with breathing anisotropy}

\author{C\'ecile Repellin$^1$}
\author{Yin-Chen He$^2$}
\author{Frank Pollmann$^3$}
\affiliation{$^1$ Max-Planck-Institut f\"ur Physik komplexer Systeme, 01187 Dresden, Germany \\
$^2$ Department of Physics, Harvard University, Cambridge, Massachusetts, 02138, USA \\
$^3$ Technische Universit\"at M\"unchen, Physics Department T42, 85747 Garching, Germany}

\date{\today}

\begin{abstract}
We numerically study the spin-$1/2$ breathing kagome lattice. In this variation of the kagome Heisenberg antiferromagnet, the spins belonging to upward and downward facing triangles have different coupling strengths. Using the density matrix renormalization group (DMRG) method and exact diagonalization, we show that the kagome antiferromagnet spin liquid is extremely robust to this anisotropy. Materials featuring this anisotropy -- and especially the recently studied vanadium compound $[{\mathrm{NH}}_{4}{]}_{2}[{\mathbf{C}}_{7}{\mathbf{H}}_{14}\mathbf{N}][{\mathbf{V}}_{7}{\mathbf{O}}_{6}{\mathbf{F}}_{18}]$ (DQVOF) -- may thus be very good candidates to realize the much studied kagome spin liquid. Further, we closely examine the limit of strong breathing anisotropy and find indications of a transition to a nematic phase.
\end{abstract}

\maketitle

\section{Introduction}
Quantum spin liquids~\cite{ANDERSON1973153} (QSLs) are strongly correlated phases which cannot be characterized by a spontaneous symmetry breaking at zero temperature. These highly entangled states feature exotic fractionalized spin excitations~\cite{Wen1990, 2016arXiv160103742S}, and have thus received a lot of theoretical and experimental interest~\cite{ANDERSON1973153, ANDERSON1196, Lee1306, BalentsNature2010}. Frustrated magnets, where the strong quantum fluctuations prevent the magnetic ordering of the spins, are a fertile ground to realize such states. While there is substantial evidence that various candidate materials might present spin liquid-like behavior, the absence of a local order parameter makes it very hard to identify QSLs experimentally. To overcome these difficulties, a theoretical understanding of the underlying models is essential.

The spin $S = 1/2$ kagome lattice with nearest-neighbor Heisenberg interactions (KAFM) is one of the prime candidates to realize a QSL. Despite intense research during the last few decades~\cite{2011Sci...332.1173Y, ReadSachdev-PhysRevLett.66.1773, Yang-PhysRevLett.70.2641, Hastings-PhysRevB.63.014413, Elser-PhysRevLett.62.2405, Marston-JAppPhys1991, nikolic-PhysRevB.68.214415, singh-PhysRevB.76.180407, evenbly-PhysRevLett.104.187203, ran-PhysRevLett.98.117205, iqbal-PhysRevB.87.060405, depenbrock-PhysRevLett.109.067201, Jiang-2012NatPh...8..902J}, the KAFM is still not fully understood.
While there is a consensus that the ground state is a QSL, it is still not fully agreed upon whether it is gapped or gapless.
A gapped $\mathbb{Z}_2$ spin liquid~\cite{depenbrock-PhysRevLett.109.067201, Jiang-2012NatPh...8..902J} is supported by early DMRG results, but a recent DMRG study brought a new perspective on the problem by unveiling signatures of Dirac cones in the transfer matrix spectrum of the ground state~\cite{2016arXiv161106238H}. These signatures are compatible with earlier variational calculations showing a competitive energy for a gapless $U(1)$ Dirac spin liquid~\cite{ran-PhysRevLett.98.117205}. On the experimental side, there is a growing number of candidate materials, such as herbertsmithite~\cite{PhysRevLett.98.107204}, volborthite~\cite{doi:10.1143/JPSJ.78.043704, doi:10.1143/JPSJ.70.3377} or vesignieite~\cite{doi:10.1143/JPSJ.78.033701}. Various perturbations modify the KAFM in these materials, such as impurities, as well as spatial or spin anisotropy or Dzyaloshinskii-Moriya interactions.

Recently, a vanadium-based compound -- $[{\mathrm{NH}}_{4}{]}_{2}[{\mathbf{C}}_{7}{\mathbf{H}}_{14}\mathbf{N}][{\mathbf{V}}_{7}{\mathbf{O}}_{6}{\mathbf{F}}_{18}]$ or diammonium quinuclidinium vanadium(III,IV) oxyfluoride (DQVOF)  -- has received significant attention. 
Experiments~\cite{natureChem-Aidoudi2011, PhysRevLett.110.207208} have shown the absence of magnetic ordering at low temperature, as well as signatures of a gapless spin liquid~\cite{PhysRevLett.110.207208, PhysRevLett.118.237203}. In this material, the $d^1$ ions $V^{4+}$ -- instead of the more usual $d^9$ $Cu^{2+}$ -- form an array of kagome layers separated by $V^{3+}$ triangular interlayers. This offers alternative properties which make DQVOF a promising kagome QSL candidate.
The use of $V^{4+}$ indeed avoids the intrinsic stoichiometric defects due to the substitution of $Zn^{2+}$ ions in herbertsmithite, and Ref.~\onlinecite{natureChem-Aidoudi2011} reported the absence of any $V^{3+}$ defects.
Thanks to relatively small interactions between the kagome and intermediary layers~\cite{natureChem-Aidoudi2011, PhysRevLett.110.207208}, the two-dimensional approximation is also better verified than in herbertsmithite. Moreover, this $V^{4+}$ material may circumvent the important Jahn-Teller distortions that lower the symmetry of $Cu^{2+}$-based kagome systems~\cite{natureChem-Aidoudi2011}. While equilateral, the triangles of the kagome lattice of DQVOF differ in size, resulting in an anisotropy of the exchange coupling between spins belonging to the upward and downward facing triangles, or breathing anisotropy. This raises an important question: are the signatures observed in DQVOF a result of the breathing anisotropy, or are they generic features of a kagome spin liquid?

Beyond the modeling of realistic materials, the numerical study of the breathing kagome model is also motivated by more theoretical aspects.
Recently, various microscopic terms have been considered to drive the KAFM to one of its neighboring phases. For example, sufficiently strong longer range interactions cause the system to spontaneously break time-reversal invariance and form a chiral spin liquid~\cite{PhysRevLett.112.137202, gongCSL-scReports}. Easy-plane and easy-axis anisotropy have also been shown to adiabatically connect the KAFM to the corresponding strong anisotropic limits~\cite{he-PhysRevLett.114.037201, lauchli-2015arXiv150404380L,2017arXiv170304659C}. 
These approaches have brought important insights into the KAFM problem in recent years~\cite{he-2015arXiv151205381H}.
We study the phase diagram of the breathing kagome model in a similar spirit. 
This deviation from the ideal kagome model is particularly interesting since it does not change the degeneracy of the classical ground state. 
The very anisotropic limit of this model has been considered in previous works~\cite{PhysRevLett.81.2356, PhysRevB.71.214413, PhysRevB.52.1133, PhysRevB.62.9484, PhysRevLett.93.187205, PhysRevB.70.104424, Schaffer2017} as a strong coupling approach to the isotropic kagome model.
However, the nature of the ground state in the strong breathing anisotropy limit remains unknown.

In this paper, we study the phase diagram of the kagome model with breathing anisotropy. Using DMRG on infinitely long cylinders and exact diagonalization of finite toroidal clusters, we numerically show that the much studied kagome spin liquid is stable in a regime extending far beyond the value of the anisotropy measured in DQVOF. 
The robustness of the kagome spin liquid is supported by several numerical signatures, including low-energy spectra, entanglement entropy, correlation length and persisting signatures of Dirac cones in the transfer matrix spectrum.
We also explore the very strongly anisotropic regime. Our finite-size results indicate a phase transition in this regime, leading to a nematic phase that preserves the translation but breaks the discrete lattice rotational symmetry of the system. We show that this state is adiabatically connected to the ground state of the first-order perturbation theory of the trimerized model~\cite{PhysRevLett.81.2356}. We also discuss the finite-size extrapolation of these results. 

The remainder of the paper is organized as follows. In Sec.~\ref{sec:theory}, we give the details of the model and review the known physics of the strongly anisotropic limit. In Sec.~\ref{sec:stability}, we show the large robustness of the kagome spin liquid with breathing anisotropy by confronting the results of iDMRG and exact diagonalization.
In Sec.~\ref{sec:strong anisotropy}, we focus on the strongly anisotropic limit. We identify a nematic state and discuss its stability in the thermodynamic limit.

\section{Description of the model and limiting cases}
\label{sec:theory}

In this section, we present our model and review the isotropic and strongly anisotropic cases.

\subsection{The kagome model}

\label{sec:model}

\begin{figure}
\begin{center}
\includegraphics[width = 0.98\linewidth]{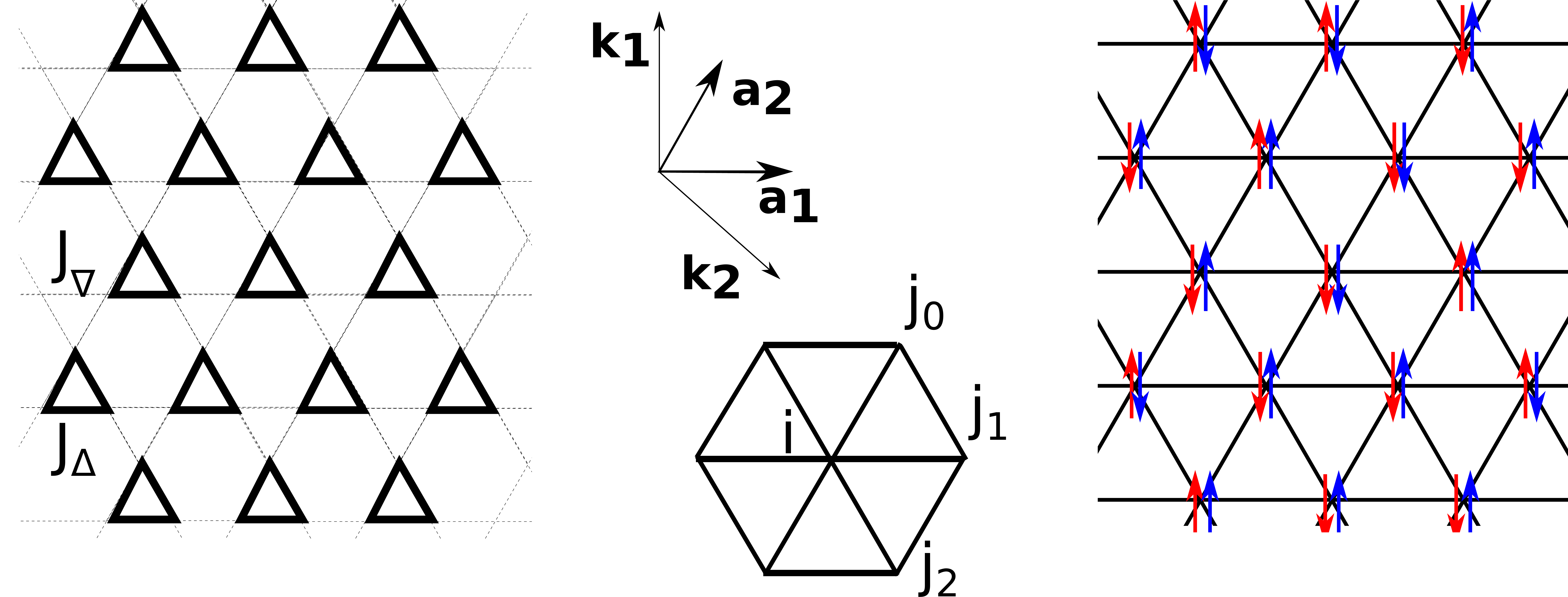}
\end{center}
\caption{Sketch of the breathing kagome lattice model. The \emph{left panel} shows the exact model Eq.~\eqref{eq:Hamiltonian}: neighboring spins belonging to an upward (resp. downward) facing triangle interact via a Heisenberg interaction of strength $J_{\upT} = 1$ (resp. $J_{\dwT} < 1$). The weak and strong bonds of the Hamiltonian are represented with lines of different thickness. The Bravais lattice $(\mathbf{a_1}, \mathbf{a_2})$ and reciprocal lattice $(\mathbf{k_1}, \mathbf{k_2})$ vectors are also represented.
The \emph{right panel} sketches the kagome lattice in the strongly trimerized limit $J_{\dwT}/J_{\upT} \ll 1$. In this limit, each set of three spins forming an upward triangle can be locally projected onto its $S = 1/2$ subspace leading to Eq.~\eqref{eq: triangle hamiltonian}. This projection maps the kagome lattice model to an effective triangular lattice where each site has a pseudospin (red) and a spin (blue) degree of freedom. In the effective model, the interactions depend on the direction of the nearest-neighbor bond $\langle i,j_{0,1,2}\rangle $
}
\label{fig:kagome}
\end{figure}

We study the spin-$1/2$ kagome lattice model with antiferromagnetic Heisenberg nearest-neighbor interactions and breathing anisotropy. The difference of size between the up and down triangles is accounted for by coupling constants of different strengths $J_{\upT}$ and $J_{\dwT}$ (see Fig.~\ref{fig:kagome}). The Hamiltonian reads
\begin{equation}
\label{eq:Hamiltonian}
H = J_{\upT}\sum_{\langle i,j\rangle  \in \upT} \mbf{S}_i \cdot \mbf{S}_j + J_{\dwT} \sum_{\langle i,j\rangle  \in \dwT} \mbf{S}_i \cdot \mbf{S}_j
\end{equation}
where $\langle i,j\rangle $ represents neighboring sites belonging to the upward ($\langle i,j\rangle  \in \upT$) or downward ($\langle i,j\rangle  \in \dwT$) facing triangles.  Without loss of generality, we set $J_{\dwT} \leq J_{\upT}$. 
The isotropic kagome lattice has the symmetry of the dihedral group $D_6$. When $J_{\dwT} \neq J_{\upT}$, inversion symmetry is broken, but rotations of order $3$ and reflections are preserved, such that the symmetry group of the lattice is the dihedral group $D_3$.

\subsection{The trimerized kagome model}
\label{sec:effective model}

Here, we focus on the limit $J_{\dwT} / J_{\upT} \ll 1$ of the breathing kagome model Eq.~\eqref{eq:Hamiltonian}, a limit known as the trimerized kagome model.
Several variants~\cite{PhysRevB.52.1133, PhysRevLett.81.2356, PhysRevB.71.214413, PhysRevB.62.9484, PhysRevLett.93.187205, Mambrini2000} of this limit have been considered in an effort to gain insight on the kagome antiferromagnet using a strong-coupling approach. We limit ourselves to the first-order perturbation theory in $J_{\dwT} / J_{\upT}$ of the trimerized kagome model, which corresponds to a frustrated spin-orbital model on the triangular lattice.

Let us review this first-order strong-coupling expansion~\cite{PhysRevLett.81.2356, PhysRevB.52.1133, PhysRevB.62.9484, PhysRevB.71.214413}. One starts ($J_{\dwT} = 0$) with a lattice of uncoupled $\upT$ triangles. A single triangle is described by the following Hamiltonian
\begin{eqnarray}
\label{eq: single triangle}
h_{\upT} & = &\frac{J_{\upT}}{2}\left[\left(\sum_{i = 1}^3 \mathbf{S}_i \right)^2 - \sum_{i = 1}^3 \mathbf{S}_i^2\right] \\ \nonumber
         & = & \frac{J_{\upT}}{2}\left[S^2 - \frac{9}{4}\right]
\end{eqnarray}
where $S = \sum_{i = 1}^3 \mathbf{S}_i$ is the total spin.
The spectrum of one isolated triangle is thus made of two $S = 1/2$ doublets with energy $-3J_{\upT}/4$, and one $S = 3/2$ quadruplet with energy $J_{\upT}/4$.
We note $\ket{{}^{s}_{\pm}}$ the four $S = 1/2$ eigenstates of Eq.~\eqref{eq: single triangle}, where the upper index $s = \uparrow, \downarrow$ corresponds to the value of the magnetization $S_z = \pm 1/2$, while the lower index $\pm$ is a pseudospin index. 
The state $\ket{{}^{s}_{\pm}}$ may be chosen to be eigenstate of the chirality operator on a triangle. However, we find the following orthonormal basis choice~\cite{PhysRevB.71.214413, PhysRevLett.93.187205} to be more convenient since it does not involve any complex numbers:
\begin{eqnarray}
\label{eq: basis}
\ket{{}^{s}_{+}} & = & \frac{1}{\sqrt{2}}\left[\ket{s s -s} - \ket{s -s s}\right] \\ \nonumber
\ket{{}^{s}_{-}} & = & \frac{1}{\sqrt{6}}\left[\ket{s s -s} + \ket{s -s s} -2 \ket{-s s s} \right] 
\end{eqnarray}

For a kagome lattice with $N_s$ spins, the many-body basis $ \{\upT\}$ is obtained by considering all $4^{N_s/3}$ possible product states formed using the $\ket{{}^{s}_{\pm}}$ single-triangle basis. 
The first-order perturbation theory thus yields:
\begin{equation}
H_\mathrm{eff} = {\cal P}_{\{\upT\}} \ H \ {\cal P}_{\{\upT\}}
\end{equation}
i.e. the projection of the original Hamiltonian Eq.~\eqref{eq:Hamiltonian} onto the basis $ \{\upT\}$. It yields an effective Hamiltonian $H_\mathrm{eff}$ on the triangular lattice with nearest-neighbor interactions entangling spin and pseudospin degrees of freedom.
Note that the form of the interaction is determined by the direction of the link connecting the two neighbors. The effective Hamiltonian reads
\begin{eqnarray}
\label{eq: triangle hamiltonian}
H_\mathrm{eff} & = & -\frac{N_s}{4}\\ \nonumber
        & + & J_{\dwT}/J_{\upT}\ \ \sum_{i}\biggl[ \left(\mbf{S}_i \otimes I^C_i \right) \cdot  \left(\mbf{S}_{j_0} \otimes I^A_{j_0} \right) \\ \nonumber
        &   & \ \ \ \ \ \ \ \  + \ \left(\mbf{S}_i \otimes I^B_i \right) \cdot  \left(\mbf{S}_{j_1} \otimes I^A_{j_1} \right) \\ \nonumber
        &   & \ \ \ \ \ \ \ \ + \ \left(\mbf{S}_i \otimes I^B_i \right) \cdot  \left(\mbf{S}_{j_2} \otimes I^C_{j_2} \right) \biggr]
        \end{eqnarray}
where each value of $i$ represents one site on the triangular lattice and $\mbf{S}_i$ is a spin-$1/2$ operator. $j_0$, $j_1$ and $j_2$ are three of the six nearest neighbors of site $i$ whose position is specified in Fig.~\ref{fig:kagome}. $I^A_i$, $I^B_i$ and $I^C_i$ are operators acting on the pseudospin degree of freedom at site $i$. Their action can be represented by the following matrices
\begin{eqnarray}
I^A & = & \left(\begin{BMAT}(b){cc}{cc}
1 & 0 \\
0 & -1/3
 \end{BMAT} \right) \\
I^B & = & \left(\begin{BMAT}(b){cc}{cc}
0 & 1/\sqrt 3 \\
1/\sqrt 3 & 2/3
 \end{BMAT} \right) \\
I^C & = & \left(\begin{BMAT}(b){cc}{cc}
0 & -1/\sqrt 3 \\
-1/\sqrt 3 & 2/3
 \end{BMAT} \right)
\end{eqnarray}

The pseudospin degree of freedom may be seen as an orbital degree of freedom, such that $H_\mathrm{eff}$ describes a Kugel-Khomskii~\cite{Kugel’:1982} interaction on the triangular lattice.  The entanglement of spin and orbital degrees of freedom in these models usually enhances the quantum fluctuations and can give rise to exotic phases of matter. On the triangular lattice, a large variety of phases has been predicted~\cite{penc2003quantum, PhysRevLett.81.3527, PhysRevB.70.014428} in spin-orbital models, including quantum spin liquids. Note, however, that these results were obtained after enforcing important symmetry constraints which do not apply to our model. While the spin degree of freedom is associated with an $SU(2)$ symmetry inherited from the original spin model, the orbital degree of freedom in $H_\mathrm{eff}$ is not associated with an $SU(2)$ symmetry or any of its subgroups. 

The nature of the ground state of the trimerized kagome model has been investigated in previous works in various ways. The nature of the ground state of $H_\mathrm{eff}$ was discussed in Ref.~\onlinecite{PhysRevB.71.214413} based on a quantum dimer model whose large resonances were argued to lead to the formation of a spin liquid. To our knowledge, the only direct numerical study of $H_\mathrm{eff}$ was performed in the one-dimensional limit of a chain of coupled triangles~\cite{PhysRevB.62.9484}.
While most of our results are based on the numerical simulation of the full model Eq.~\eqref{eq:Hamiltonian}, we will also use this effective trimerized model and relate its ground state to the ground state of the kagome model.

\section{Stability of the kagome spin liquid under breathing anisotropy}
\label{sec:stability}

In this section, we examine the properties of the ground state of the kagome model with breathing anisotropy described by Eq.~\eqref{eq:Hamiltonian}. The DMRG algorithm has provided important insights into the nature of the ground state of the KAFM and some of its variants in the past and is thus a method of choice here. This method allows to efficiently obtain the ground state of our microscopic model on long cylinders. We also use exact diagonalization  to obtain the low-energy spectrum in geometries preserving all the symmetries of the model. Finally, we extract the transfer matrix spectrum at various values of the breathing anisotropy, and show that the recent observation~\cite{2016arXiv161106238H} of Dirac cone signatures in the isotropic model stays valid even for rather strong breathing anisotropy.

\subsection{Properties of the ground state on an infinite cylinder: iDMRG results}
\label{sec:stability DMRG}
We employ the infinite version of the DMRG algorithm~\cite{PhysRevLett.69.2863, RevModPhys.77.259, 2011AnPhy.326...96S, 2dDMRGStoudenmireWhite, PhysRevB.87.235106} (iDMRG) to compute the ground state of Eq.~\eqref{eq:Hamiltonian} on the cylinder geometry. The cylinder axis is along the $x$-direction, and the $y$ axis coincides with one of the lattice axes. This geometry is commonly called $\mathrm{YC2n}$, where $2n$ is the number of sites along the circumference. Cylinders which connect periodically across the cylinder with an additional two site shift along the $x$ direction are denoted $\mathrm{YC2n}-2$. DMRG is a one-dimensional method that can also be used to find the ground state of two-dimensional systems by considering a chain forming a "snake" around the cylinder circumference. A translationally invariant state on the $\mathrm{YC2n}$ cylinder can be described by a uniform matrix product state (MPS) using $3n$ matrices, one for each site of the $1D$ unit cell. In contrast, the size of the $\mathrm{YC2n}-2$ unit cell is $6$ independently of $n$. Since the numerical cost of the iDMRG algorithm 
varies linearly with the size of the unit cell, the $\mathrm{YC2n}-2$ geometry offers a significant computational advantage at large circumference $n$. We thus computed the ground state of Hamiltonian~\eqref{eq:Hamiltonian} in the $\mathrm{YC8}$, $\mathrm{YC8}-2$, $\mathrm{YC10}-2$ and $\mathrm{YC12}-2$ geometries.

\begin{figure}
\includegraphics[width=0.49\linewidth]{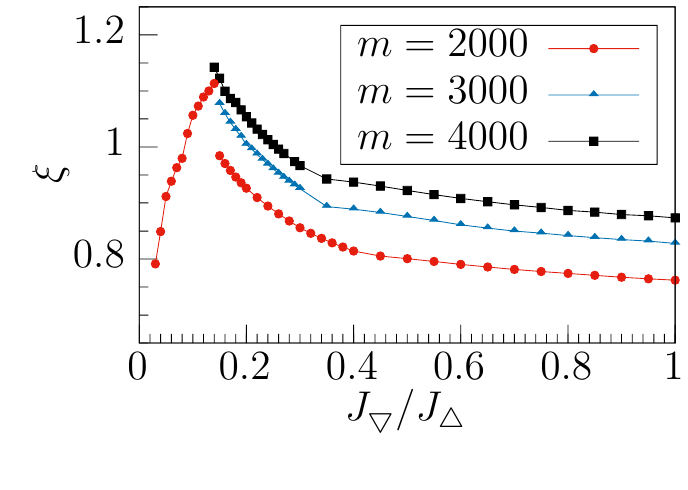}
\includegraphics[width=0.49\linewidth]{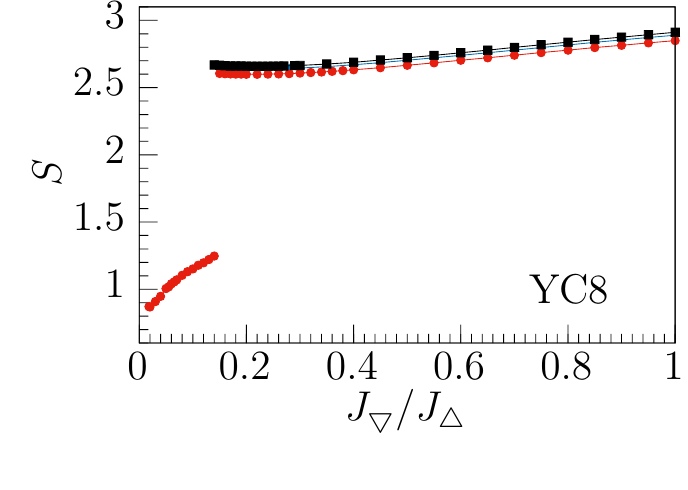}
\caption{Evolution of the correlation length $\xi$ and entanglement entropy $S$ of the ground state of the breathing kagome model Eq.~\eqref{eq:Hamiltonian} on an infinitely long $\mathrm{YC8}$ cylinder. Up to $m = 4000$ states were kept in the $S_z$ conserving DMRG simulation. 
For small $J_\dwT/J_\upT$, the results show  negligible dependence on the bond dimension $m$.
}
\label{fig:finite anisotropy}
\end{figure}

In the isotropic limit $J_{\dwT} / J_{\upT} = 1$, the ground state of the Hamiltonian Eq.~\eqref{eq:Hamiltonian} was identified as a time-reversal symmetric spin liquid in earlier works~\cite{2011Sci...332.1173Y}. As noted in Refs.~\onlinecite{2011Sci...332.1173Y, depenbrock-PhysRevLett.109.067201}, DMRG simulations converge to a ground state with short range correlations on the cylinder geometry. We find that this behavior persists when a finite, and even relatively large amount of anisotropy is introduced in the model. Fig.~\ref{fig:finite anisotropy} shows the evolution of various observables on the $\mathrm{YC8}$ cylinder for several values of the anisotropy parameter $J_{\dwT} / J_{\upT}$, keeping up to $m = 4000$ states in the iDMRG simulation. 
The numerical results in the other cylinder geometries ($\mathrm{YC8}-2$, $\mathrm{YC10}-2$ and $\mathrm{YC12}-2$) are given in the Appendix~\ref{app:DMRG}.
The correlation length is extracted from the two leading eigenvalues of the transfer matrix and thus conveniently obtained in our MPS formalism.
The entanglement entropy is obtained for a bipartition of the system cutting the cylinder along the $y$ axis.
Both the correlation length and the entanglement entropy do not show any sign of a phase transition over a very wide interval of the anisotropy parameter ($J_{\dwT} / J_{\upT} \geq 0.14$ in the $\mathrm{YC8}$ geometry, and at least up to $J_{\dwT} / J_{\upT} \simeq 0.2$ in all the considered cylinder geometries). 

As can be seen in Fig.~\ref{fig:finite anisotropy}, all observables show a sharp transition at a value $J_{\dwT} / J_{\upT} \simeq 0.14$. This transition also exists in the other geometries, albeit at different values of $J_{\dwT} / J_{\upT}$. We will discuss this phase transition in a later paragraph (see Sec.~\ref{sec:strong anisotropy}).

\subsection{Low energy spectra from exact diagonalization}
\label{sec:ED}

\begin{figure}
\begin{center}
\includegraphics[width=0.9\linewidth]{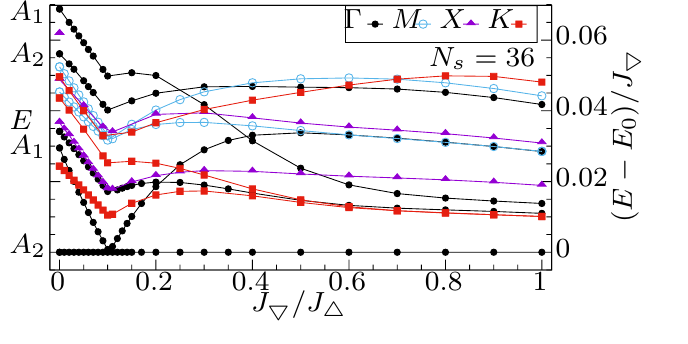}
\end{center}
\caption{Low energy spectrum of a kagome cluster with periodic boundary conditions and $N_s = 36$ spins.
The states are labeled by their momentum sector and by the Mulliken symbol for the irreducible representation of $D_3$ in the $\Gamma$ sector. The angular point at $J_{\dwT}/J_{\upT} \simeq 0.10$ is due to a crossing between the ground states of the $\Gamma_{A_1}$ and $\Gamma_{A_2}$ sectors. A crossing occurs at an almost identical value of $J_{\dwT}/J_{\upT}$ in the $M$ sector. In the $K$ sector, the breaking of inversion symmetry for $J_{\dwT} \neq J_{\upT}$ results in the lifting of the associated degeneracy.}
\label{fig:ed finite anisotropy}
\end{figure}

Additionally to our iDMRG analysis, we studied the model Eq.~\eqref{eq:Hamiltonian} using exact diagonalization. While only smaller system sizes can be reached with this technique, it is a method of choice to obtain the low-energy spectrum beyond the ground state. It also facilitates the exploration of systems preserving all spatial symmetries of the model.

We consider a finite kagome cluster with periodic boundary conditions in both directions. We restrict ourselves to clusters with an even number of spins $N_s$ to allow the formation of spin singlets in the system.
As noted in Sec.~\ref{sec:theory}, the kagome lattice with breathing anisotropy has the symmetry of the dihedral group $D_3$. The three smallest toroidal clusters which preserve this symmetry are made of $N_s = 12, 36$ and $48$ spins. The $N_s = 48$ cluster is beyond our reach (its ground state across all sectors was recently obtained~\cite{2016arXiv161106990L} in the isotropic case -- which preserves additional symmetries -- at the cost of tremendous computational effort). This system can, however, be studied in the strongly anisotropic limit using the trimerized model Eq.~\eqref{eq: triangle hamiltonian} thanks to the dimensional reduction due to the projection onto local trimers. We will take advantage of this property in Sec.~\ref{sec:strong anisotropy}.

We obtained the low-energy spectrum for clusters up to $N_s = 36$ with several values of the breathing anisotropy (see Fig.~\ref{fig:ed finite anisotropy} for $N_s = 36$ and Appendix~\ref{app:ED} for the other sizes).
The singlet-triplet gap remains large compared to the typical energy difference between successive energy levels, and we thus focus on the part of the spectrum with only singlets. The states are labeled by their momentum sector, as well as their eigenvalues under the $D_3$ operations when this symmetry is present. For all system sizes, the (normalized) low-energy spectrum is remarkably unaffected by an important change of the breathing anisotropy. Indeed, the relative positions of the first few eigenstates is unchanged for $ 0.55 \leq J_{\dwT}/J_{\upT} \leq 1$ for all system sizes.
More specifically, the smallest sizes ($N_s \leq 24$) do no not show any gap closing above the ground state for any value of $J_{\dwT}/J_{\upT}$ such that the ground states in the isotropic and strongly anisotropic limits are adiabatically connected. However in the $N_s = 30$ and $36$ systems, the respective ground states in these two limits are characterized by different quantum numbers. The level crossing between the two states takes place respectively at $J_{\dwT}/J_{\upT} \simeq 0.4$ and $J_{\dwT}/J_{\upT} \simeq 0.1$. The $N_s = 36$ system may be the most meaningful, since it preserves the full symmetry of the lattice. In this system, this transition corresponds to a crossing between the ground states of the $\Gamma_{A_1}$ and $\Gamma_{A_2}$ symmetry sectors (see Fig.~\ref{fig:ed finite anisotropy}).

The stability of the low-energy spectrum in all finite-size clusters confirms the robustness of the kagome spin liquid, in agreement with the iDMRG results  of the previous paragraph. The largest clusters feature level crossings at very small values of $J_{\dwT}/J_{\upT}$. The similarity of these crossings with the one observed in the infinite cylinders -- and the possibility that they might reveal the same phase transition -- will be discussed in Sec.~\ref{sec:strong anisotropy}.

\subsection{Signatures of Dirac cones using the transfer matrix spectrum}
Using iDMRG, it is possible to obtain the momentum-resolved spectrum of correlation lengths, which is related to the excitation spectrum~\cite{1367-2630-17-5-053002}. Ref.~\onlinecite{2016arXiv161106238H} has recently provided insights on the nature of the kagome ground state using this method at the isotropic point $J_{\dwT} = J_{\upT}$. This study has reported signatures of Dirac cones, giving additional credibility to the hypothesis that the kagome ground state could be a Dirac spin liquid. The discrepancy between this observation and previous reports of a finite gap in DMRG studies was interpreted as a consequence of the wrapping of the system on a narrow cylinder. In this geometry, only few momentum values are allowed along the $y$ direction.
According to Ref.~\onlinecite{2016arXiv161106238H}, the putative Dirac points fall on forbidden momenta leading to the opening of a finite gap to minimize the energy of the ground state.

\begin{figure}
\includegraphics[width = 0.98\linewidth]{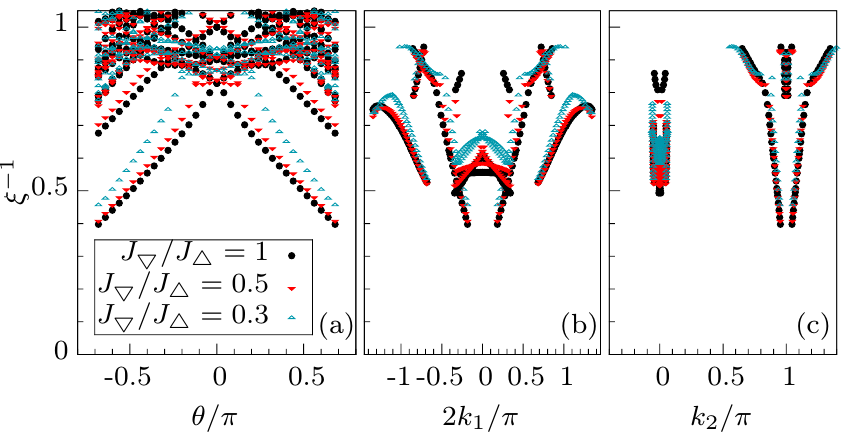}
\caption{(a) Evolution of the inverse of the correlation length $\xi$ in the $\mathrm{YC8}-2$ infinite cylinder upon adiabatic insertion of a flux $\theta$ for various values of the anisotropy parameter. $m = 4000$ states were kept in the DMRG simulation. (b, c) The momentum-resolved correlation length spectrum was extracted following the procedure outlined in Ref.~\onlinecite{2016arXiv161106238H}.}
\label{fig:corr length mom}
\end{figure}

We apply the same approach to the breathing kagome model. The momentum-resolved spectrum of correlation lengths is obtained from the leading eigenvalues of the transfer matrix, which is obtained from the MPS formulation of the ground state.
Adiabatically inserting a flux $\theta$ along the cylinder axis is equivalent to twisting the boundary conditions of the cylinder, and reveals the properties of the system at various momenta. We use the $\mathrm{YC8-2}$ geometry which provides a clear Dirac cone signature in Ref.~\onlinecite{2016arXiv161106238H}. Our results are represented in Fig.~\ref{fig:corr length mom} for $J_{\dwT}/J_{\upT} = 1$ (reproducing the results of Ref.~\onlinecite{2016arXiv161106238H}), $0.5$, and $0.3$. We show the inverted correlation length as a function of the twisting angle $\theta$, and as function of the two momentum components $(k_1, k_2)$ (defined in Fig.~\ref{fig:kagome}), which are extracted from the complex phase of the transfer matrix eigenvalues. The essential features stay the same throughout the phase. This includes similar Dirac cone signatures in the momentum-resolved spectrum. 

We note that the data near the Dirac point $(2k_1, k_2)=(0, \pi)$ is missing in Fig.~\ref{fig:corr length mom}, which is because we cannot adiabatically insert flux  when $\theta\sim \pi$. 
Physically it comes from the  instability of the Dirac spin liquid on a quasi-1D system~\cite{2016arXiv161106238H}. 
One can in principle stabilize the Dirac spin liquid when $\theta\sim\pi$ by adding further neighbor interactions ($J_2$) and by doing the adiabatic flux insertion with a smaller flux increment (see Ref.~\onlinecite{2016arXiv161106238H} for further details).
We did not follow this direction here since the missing points do not obfuscate the main conclusion: both the Dirac cone structures and the instability of the spin liquid near $\theta \sim \pi$ persist from $J_\dwT/J_\upT= 1$ to $J_\dwT/J_\upT\sim 0.3$.
This analysis -- based on a probe very sensitive to any change of the ground state nature -- confirms the remarkable robustness of the kagome spin liquid with regards to the breathing anisotropy.

\vspace{12pt}

We have explored the ground state properties of the kagome lattice model with breathing anisotropy. Several signatures -- extracted from iDMRG and exact diagonalization -- reveal a robust kagome spin liquid. The iDMRG study shows the existence of a phase transition at small $J_{\dwT}/J_{\upT}$, beyond which the ground state spontaneously breaks the discrete lattice rotational symmetry of the model. A transition is also observed in the largest finite clusters studied with exact diagonalization. In the following section, we focus on the strongly anisotropic regime of the model. We analyze the nature of the ground state in this regime and discuss the extrapolation of our results to the thermodynamic limit, in particular the existence of a phase transition at finite, but strong breathing anisotropy.

\section{Probing the strongly anisotropic limit}
\label{sec:strong anisotropy}

We now focus on the strongly anisotropic regime of Eq.~\eqref{eq:Hamiltonian} ($J_{\dwT}/J_{\upT} < 0.2$). In most system geometries, we have observed a phase transition in this regime. In this section, we show that the ground state beyond the transition displays signatures of nematic order. We discuss the possible finite-size extrapolation of this result. 

\subsection{Signatures of nematic order in the strongly anisotropic limit}

\begin{figure}
\begin{center}
\includegraphics[width = 0.53\linewidth]{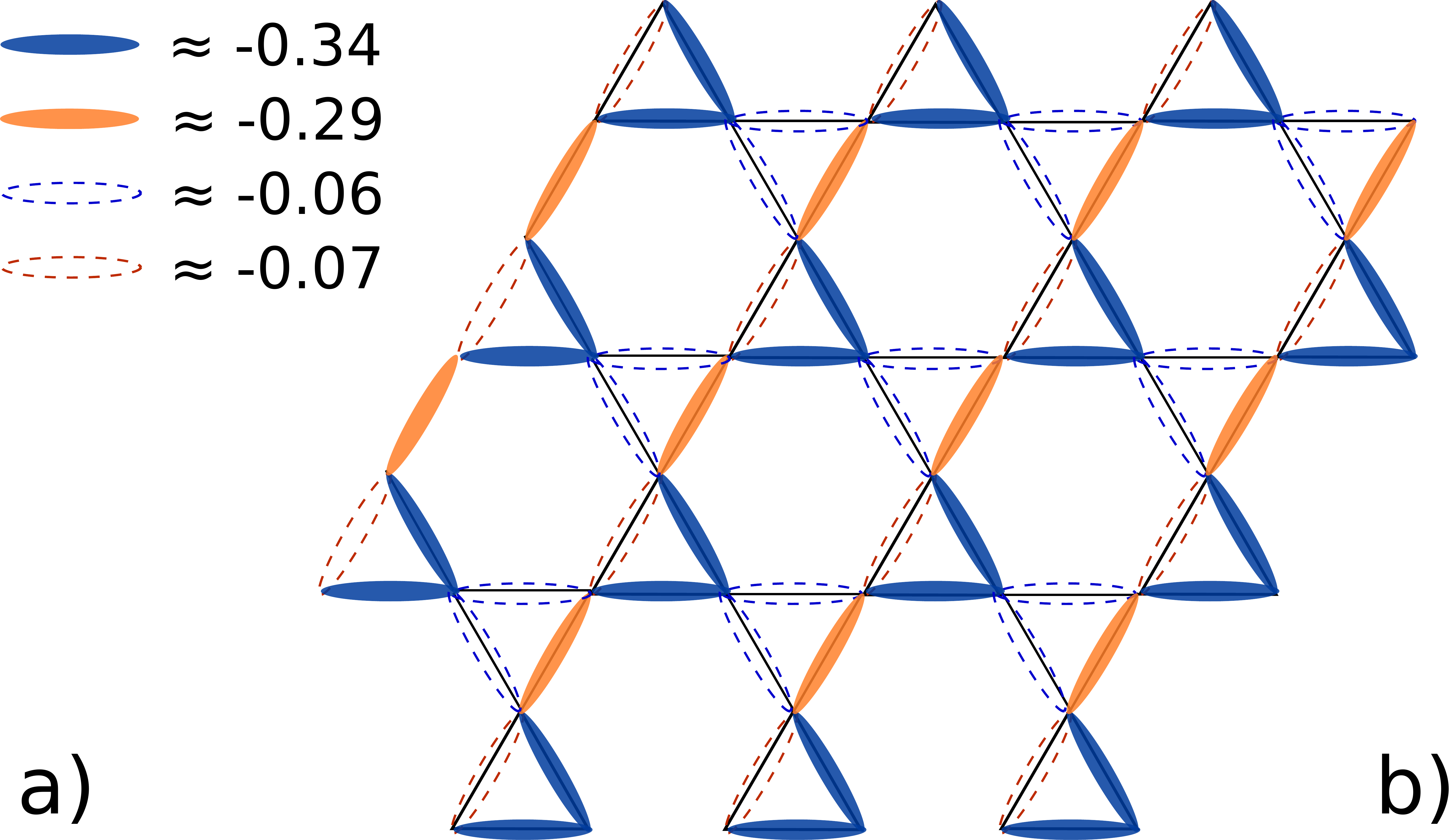}
\includegraphics[width=0.44\linewidth]{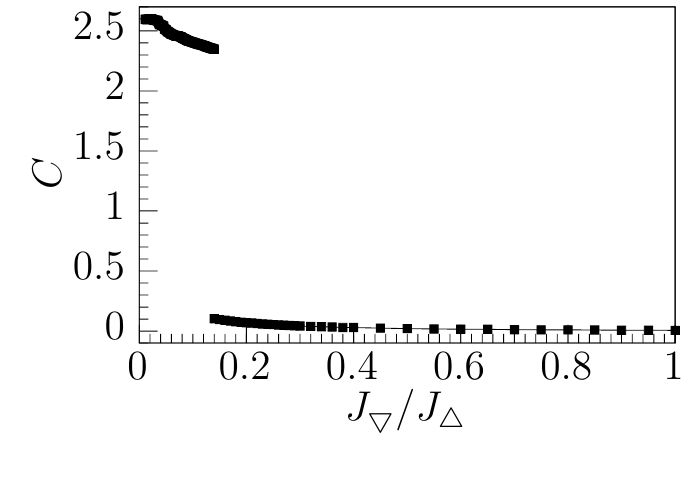}
\end{center}
\caption{(a) Neighboring spin-spin correlations in the ground state of the breathing kagome model in the limit $J_{\dwT}/J_{\upT} \ll 1$. The numerical values were extracted from the ground state of $H_\mathrm{eff}$ on the $\mathrm{YC12}$ geometry, obtained using iDMRG. (b) Evolution of the nematic order parameter $C$ \eqref{eq:nematic OP} of the ground state of the breathing kagome model Eq.~\eqref{eq:Hamiltonian} on an infinitely long $\mathrm{YC8}$ cylinder. The results show  negligible dependence on the bond dimension $m$, which was set to $m = 4000$ here.}
\label{fig:nematic}
\end{figure}

We analyze the properties of the $J_{\dwT} < J_{\dwT}^c$ ground state using iDMRG and exact diagonalization.
The main results are summarized here, and additional details can be found in Appendix~\ref{app:DMRG},~\ref{app:ED nematic} and ~\ref{app:corr}.

The iDMRG data reveals a spontaneous breaking of the rotational symmetry for $J_{\dwT} < J_{\dwT}^c$. The pattern of nearest-neighbor spin-spin correlations in the ground state is given in Fig.~\ref{fig:nematic}a.
We can define a nematic order parameter $C$ which measures the breaking of the threefold rotational symmetry of the lattice:
\begin{equation}
C = \frac{2}{N_s}\sum_{\langle i,j\rangle}  \gamma_{ij} \mathbf{S_i}\cdot \mathbf{S_j}
\label{eq:nematic OP}
\end{equation}
where the sum runs over all the nearest-neighbor spin pairs, and $\gamma_{ij} = e^{2\pi i \frac{l}{6}}$ is a phase defined by the direction of the $\langle i,j\rangle $ bond ($l = 0, 1, ..., 5$ clockwise around each hexagon of the kagome lattice). The phase transition is easily detected by tracking the value of $C$ as can be seen in Fig.~\ref{fig:nematic}b in the $\mathrm{YC8}$ geometry (see Appendix~\ref{app:DMRG} for the other cylinder geometries).

Note that the infinite cylinder geometry (which breaks the threefold rotational symmetry of the model) may influence the emergence of the nematic order. We thus turn to the exact diagonalization of finite clusters preserving all spatial symmetries of the model ($N_s = 12, 36, 48$). In this case, no spontaneous breaking of the $D_3$ symmetry can occur. 
Instead, we expect four low-lying states carrying the quantum numbers of the different representations of the dihedral group $D_3$: $E$ (doubly degenerate), $A_1$ and $A_2$. The strong anisotropy spectra (obtained using the trimerized model of Sec.~\ref{sec:effective model}) are compatible with this prediction, even though they show a large finite-size splitting of the fourfold quasidegeneracy (see Appendix~\ref{app:ED nematic}).
We also computed the spin-spin and dimer-dimer correlations of the ground state (see Appendix~\ref{app:corr}). The former confirm the absence of spin-spin long-range order in the anisotropic regime. The large dimer-dimer correlations form a pattern which mostly coincides with the pattern found with iDMRG.
Finally, we studied the response of the ground state to a nematic perturbation. We found that the addition of an arbitrarily small $D_3$-breaking term caused a significant response (finite value of $C$) in the strongly anisotropic regime and not for $J_{\dwT} > J_{\dwT}^c$ (see Appendix~\ref{app:ED nematic}). This confirms the large tendency to nematicity of the ground state of the kagome model with strong breathing anisotropy.

\subsection{Connecting the ground states of the models with finite and infinite (trimerized) anisotropy}
\begin{figure}
\includegraphics[width = 0.9\linewidth]{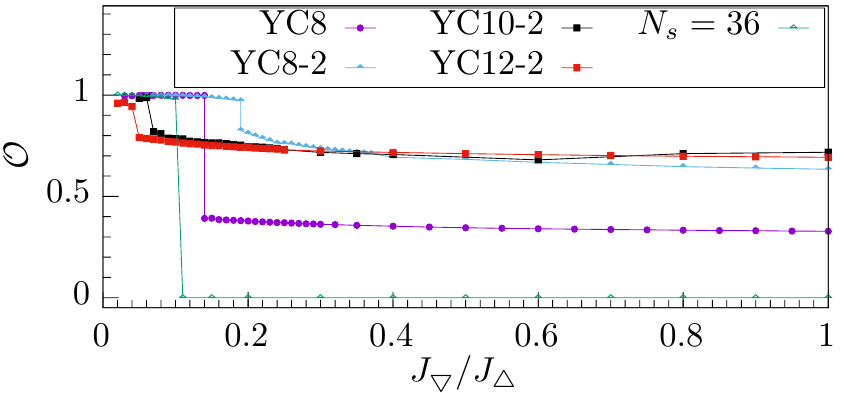}
\caption{Overlap of the ground state of the effective model Eq.~\eqref{eq: triangle hamiltonian} with the projected ground state of the breathing kagome model in the $\mathrm{YC8}$, $\mathrm{YC8}-2$, $\mathrm{YC10}-2$ and $\mathrm{YC12}-2$ geometries, and $N_s = 36$ spins toroidal cluster. On the infinite cylinder geometries, we show the overlap per site, extrapolated to $m\rightarrow \infty$ by using up to $m = 6000$ states in the iDMRG simulation.}
\label{fig: eff model overlaps}
\end{figure}

We used the effective model $H_\mathrm{eff}$ from Sec.~\ref{sec:effective model} to obtain the ground state in the strong anisotropy limit $J_{\dwT} / J_{\upT} \ll 1$. It omits the high-energy degrees of freedom irrelevant in this limit, thus allowing to reach larger sizes: the toroidal clusters up to $N_s = 48$ for exact diagonalization, and up to $\mathrm{YC12}$ cylinders with iDMRG. Consistently with the smaller sizes in the strongly anisotropic regime, we obtain a nematic ground state with short-range spin-spin correlations in the direction orthogonal to the stripes.

In the regime of small (but finite) $J_{\dwT} / J_{\upT}$, the high density of low-energy states in the exact kagome model~\eqref{eq:Hamiltonian} makes it difficult to converge to the ground state. This sets a lower bound to the accessible values of $J_{\dwT} / J_{\upT}$, leaving a blind area in the strong anisotropy regime.
To learn about this regime, we have compared the ground states of \eqref{eq:Hamiltonian} and $H_\mathrm{eff}$. To compute their overlap, we must first project the kagome ground state $\ket{\Psi_{J_{\dwT}/J_{\upT}}}$ onto the effective basis $ \{\upT\}$
\begin{equation}
{\cal P}_{\{\upT\}}  \ket{\Psi_{\mathrm{kag}}}
\end{equation}
This operation amounts to a projection of the local Hilbert space of each $\upT$ triangle onto the $S = 1/2$ basis defined in Eq.~\eqref{eq: basis}. The overlap per site $\mathcal O$ of the projected wave function with the ground state of $H_\mathrm{eff}$ is given for all values of $J_{\dwT}/J_{\upT}$ in Fig.~\ref{fig: eff model overlaps}. While $\mathcal O$ is rather small ($<0.9$) in the kagome spin liquid phase, it is very close to one ($\mathcal O > 0.99$) in the nematic phase. This very large overlap strongly suggests that the nematic ground state observed in the anisotropic kagome model for $J_{\dwT} < J_{\dwT}^c$ and the ground state of the effective model in fact correspond to the same phase. As a result, there would be at most two phases in the kagome model with breathing anisotropy, separated by a transition at $J_{\dwT} = J_{\dwT}^c$.

\subsection{Finite-size extrapolation}
The existence of a phase transition in the breathing kagome model (for a strong breathing anisotropy) is supported by numerical results from two different numerical techniques, iDMRG and exact diagonalization. From these results, we can draw a phase diagram involving the kagome spin liquid for $J_{\dwT}^c < J_{\dwT} \leq J_{\upT}$, and a nematic state for $0 < J_{\dwT} < J_{\dwT}^c$. The existence of any other phases was ruled out in the accessible sizes by using a first order expansion of the trimerized model and relating its ground state to the ground state of the kagome model with $J_{\dwT} < J_{\dwT}^c$. Note that we have found a level crossing in all the geometries where the transition was observed. While this observation evokes a first-order phase transition, it is also compatible with a continuous transition in the case of a gapless spin liquid.

An important question is the significance of these results at the thermodynamic limit. With exact diagonalization, only few system sizes are available, and even fewer ($N_s = 12, 36$ and $N_s = 48$ in the trimerized limit) have the full symmetry of the lattice. The variety of behaviors concerning the existence and position of a phase transition in these systems is thus perhaps not so surprising. DMRG lets us explore cylinders with various perimeters and gives more consistent results, but the system's geometry might include an intrinsic bias towards the observed nematic state. Moreover, the value of the anisotropy at the transition shows a slight drift to smaller $J_{\dwT}^c$ as the circumference is increased. This is especially clear when one considers the value of $J_{\dwT}^c$ in the $\mathrm{YC8}-2$, $\mathrm{YC10}-2$ and $\mathrm{YC12}-2$ geometries (respectively $0.19$, $0.07$, $0.05$ as can be seen in Fig.~\ref{fig: eff model overlaps}). Since only a few cylinder perimeters are available, it is difficult to extrapolate the value of $J_{\dwT}^c$ to the thermodynamic limit. We cannot completely rule out the possibility that it might become zero in the thermodynamic limit, and that the nematic ground state might be unstable at finite $J_{\dwT}$.

\section{Conclusion}
We have numerically studied the influence of the breathing anisotropy on the ground state of the kagome nearest-neighbor antiferromagnet. Using a variety of methods based on DMRG and exact diagonalization, we have shown that the kagome spin liquid antiferromagnet is extremely stable to the breathing anisotropy. The variety of system geometries (infinitely long cylinder and torus) and sizes is a strong argument to argue that this finding may remain qualitatively true in the thermodynamic limit and should apply to the kagome material DQVOF. Indeed, even our most conservative estimation of the stability regime ($J_{\dwT} / J_{\upT} > 0.2$) includes the value measured in this material ($J_{\dwT} / J_{\upT} \simeq 0.55$ according to a very recent estimation~\cite{PhysRevLett.118.237203}). DQVOF thus appears as an excellent candidate to realize the kagome spin liquid. Moreover, we have found signatures of Dirac cones in the ground state of the breathing kagome model which are compatible with the experimental observation of a gapless spin liquid in DQVOF. These signatures are essentially the same as the ones already observed in the isotropic model~\cite{2016arXiv161106238H}.

Beyond the connection to existing materials, we have discussed the existence of a phase transition at $J_{\dwT} / J_{\upT} < 0.2$. Such transition appears for all the cylinder geometries that we studied, as well as in the largest toroidal cluster that we were able to diagonalize ($36$ spins). 
Considering the large dependency of the breathing anisotropy at the transition $J_{\dwT}^c$ with system size (from $0.05$ for the $\mathrm{YC12}-2$ cylinder to $0.2$ for the $\mathrm{YC8}-2$ cylinder), it is hard to extrapolate these findings to the thermodynamic limit. In fact, the transition value seems to decrease when the cylinder perimeter increases in the $\mathrm{YC2n}-2$ geometry. Whether or not this value remains finite at the thermodynamic limit remains to be determined.

We have also investigated the nature of the ground state in the strongly anisotropic (trimerized) limit. On the cylinder geometry, for the accessible perimeters, the ground state has nematic order. On the torus geometry (considering only geometries which preserve the rotational symmetry of the model), the analysis of the ground state has shown a large tendency to nematicity. The quantum numbers of the lowest energy states in the $36$ and $48$ spins cluster are in agreement with this hypothesis, but the spectra show large finite size effects.
Using an effective trimerized model has allowed us to explore numerically the very anisotropic regime of the phase diagram. Interestingly, the phase diagram seems to consist of at most two phases: one connected to the ground state of the trimerized model, and one connected to the kagome spin liquid.

The nematic phase in the  strongly anisotropic (trimerized) limit preserves  the translational symmetry, the spin rotation symmetry and it has spin$-3/2$ per unit cell.
Therefore according to the  Hastings-Oshikawa-Lieb-Schultz-Mattis theorem~\cite{1961AnPhy..16..407L, PhysRevLett.84.1535, Misguich2002, 0295-5075-70-6-824}, one cannot adiabatically deform this nematic phase into a trivial product state.
This leads to interesting possibilities, for example the nematic phase could be a 2D (nematic) spin liquid.
The other possibility is that the nematic phase is made of decoupled 1D Luttinger liquids even though the original trimerized model is 2D isotropic.
 Understanding the nematic phase and the phase transition might lead to new insight on the kagome spin liquid.

\section{Acknowledgments}
We thank D. Poilblanc, A. Tsirlin, N. Regnault and M. Mambrini for insightful discussions, and especially Roderich Moessner for collaboration during the early stages of this work.
This work was supported in parts by the German Research Foundation (DFG) via SFB 1143 and Research Unit FOR 1807.
YCH is supported by a postdoctoral fellowship from the Gordon and Betty Moore Foundation, under the EPiQS initiative, GBMF4306, at Harvard University.

\bibliography{KagomeHeisenbergAnisotropy}

\begin{thebibliography}{56}%
\makeatletter
\providecommand \@ifxundefined [1]{%
 \@ifx{#1\undefined}
}%
\providecommand \@ifnum [1]{%
 \ifnum #1\expandafter \@firstoftwo
 \else \expandafter \@secondoftwo
 \fi
}%
\providecommand \@ifx [1]{%
 \ifx #1\expandafter \@firstoftwo
 \else \expandafter \@secondoftwo
 \fi
}%
\providecommand \natexlab [1]{#1}%
\providecommand \enquote  [1]{``#1''}%
\providecommand \bibnamefont  [1]{#1}%
\providecommand \bibfnamefont [1]{#1}%
\providecommand \citenamefont [1]{#1}%
\providecommand \href@noop [0]{\@secondoftwo}%
\providecommand \href [0]{\begingroup \@sanitize@url \@href}%
\providecommand \@href[1]{\@@startlink{#1}\@@href}%
\providecommand \@@href[1]{\endgroup#1\@@endlink}%
\providecommand \@sanitize@url [0]{\catcode `\\12\catcode `\$12\catcode
  `\&12\catcode `\#12\catcode `\^12\catcode `\_12\catcode `\%12\relax}%
\providecommand \@@startlink[1]{}%
\providecommand \@@endlink[0]{}%
\providecommand \url  [0]{\begingroup\@sanitize@url \@url }%
\providecommand \@url [1]{\endgroup\@href {#1}{\urlprefix }}%
\providecommand \urlprefix  [0]{URL }%
\providecommand \Eprint [0]{\href }%
\providecommand \doibase [0]{http://dx.doi.org/}%
\providecommand \selectlanguage [0]{\@gobble}%
\providecommand \bibinfo  [0]{\@secondoftwo}%
\providecommand \bibfield  [0]{\@secondoftwo}%
\providecommand \translation [1]{[#1]}%
\providecommand \BibitemOpen [0]{}%
\providecommand \bibitemStop [0]{}%
\providecommand \bibitemNoStop [0]{.\EOS\space}%
\providecommand \EOS [0]{\spacefactor3000\relax}%
\providecommand \BibitemShut  [1]{\csname bibitem#1\endcsname}%
\let\auto@bib@innerbib\@empty
\bibitem [{\citenamefont {Anderson}(1973)}]{ANDERSON1973153}%
  \BibitemOpen
  \bibfield  {author} {\bibinfo {author} {\bibfnamefont {P.}~\bibnamefont
  {Anderson}},\ }\href {\doibase
  http://dx.doi.org/10.1016/0025-5408(73)90167-0} {\bibfield  {journal}
  {\bibinfo  {journal} {Materials Research Bulletin}\ }\textbf {\bibinfo
  {volume} {8}},\ \bibinfo {pages} {153 } (\bibinfo {year} {1973})}\BibitemShut
  {NoStop}%
\bibitem [{\citenamefont {Wen}(1990)}]{Wen1990}%
  \BibitemOpen
  \bibfield  {author} {\bibinfo {author} {\bibfnamefont {X.~G.}\ \bibnamefont
  {Wen}},\ }\href {\doibase 10.1142/S0217979290000139} {\bibfield  {journal}
  {\bibinfo  {journal} {International Journal of Modern Physics B}\ }\textbf
  {\bibinfo {volume} {04}},\ \bibinfo {pages} {239} (\bibinfo {year}
  {1990})}\BibitemShut {NoStop}%
\bibitem [{\citenamefont {{Savary}}\ and\ \citenamefont
  {{Balents}}(2016)}]{2016arXiv160103742S}%
  \BibitemOpen
  \bibfield  {author} {\bibinfo {author} {\bibfnamefont {L.}~\bibnamefont
  {{Savary}}}\ and\ \bibinfo {author} {\bibfnamefont {L.}~\bibnamefont
  {{Balents}}},\ }\href@noop {} {\bibfield  {journal} {\bibinfo  {journal}
  {ArXiv e-prints}\ } (\bibinfo {year} {2016})},\ \Eprint
  {http://arxiv.org/abs/1601.03742} {arXiv:1601.03742 [cond-mat.str-el]}
  \BibitemShut {NoStop}%
\bibitem [{\citenamefont {Anderson}(1987)}]{ANDERSON1196}%
  \BibitemOpen
  \bibfield  {author} {\bibinfo {author} {\bibfnamefont {P.~W.}\ \bibnamefont
  {Anderson}},\ }\href {\doibase 10.1126/science.235.4793.1196} {\bibfield
  {journal} {\bibinfo  {journal} {Science}\ }\textbf {\bibinfo {volume}
  {235}},\ \bibinfo {pages} {1196} (\bibinfo {year} {1987})}\BibitemShut
  {NoStop}%
\bibitem [{\citenamefont {Lee}(2008)}]{Lee1306}%
  \BibitemOpen
  \bibfield  {author} {\bibinfo {author} {\bibfnamefont {P.~A.}\ \bibnamefont
  {Lee}},\ }\href {\doibase 10.1126/science.1163196} {\bibfield  {journal}
  {\bibinfo  {journal} {Science}\ }\textbf {\bibinfo {volume} {321}},\ \bibinfo
  {pages} {1306} (\bibinfo {year} {2008})}\BibitemShut {NoStop}%
\bibitem [{\citenamefont {Balents}(2010)}]{BalentsNature2010}%
  \BibitemOpen
  \bibfield  {author} {\bibinfo {author} {\bibfnamefont {L.}~\bibnamefont
  {Balents}},\ }\href {\doibase 10.1126/science.1163196} {\bibfield  {journal}
  {\bibinfo  {journal} {Nature}\ }\textbf {\bibinfo {volume} {464}},\ \bibinfo
  {pages} {199} (\bibinfo {year} {2010})}\BibitemShut {NoStop}%
\bibitem [{\citenamefont {{Yan}}\ \emph {et~al.}(2011)\citenamefont {{Yan}},
  \citenamefont {{Huse}},\ and\ \citenamefont {{White}}}]{2011Sci...332.1173Y}%
  \BibitemOpen
  \bibfield  {author} {\bibinfo {author} {\bibfnamefont {S.}~\bibnamefont
  {{Yan}}}, \bibinfo {author} {\bibfnamefont {D.~A.}\ \bibnamefont {{Huse}}}, \
  and\ \bibinfo {author} {\bibfnamefont {S.~R.}\ \bibnamefont {{White}}},\
  }\href {\doibase 10.1126/science.1201080} {\bibfield  {journal} {\bibinfo
  {journal} {Science}\ }\textbf {\bibinfo {volume} {332}},\ \bibinfo {pages}
  {1173} (\bibinfo {year} {2011})}\BibitemShut {NoStop}%
\bibitem [{\citenamefont {Read}\ and\ \citenamefont
  {Sachdev}(1991)}]{ReadSachdev-PhysRevLett.66.1773}%
  \BibitemOpen
  \bibfield  {author} {\bibinfo {author} {\bibfnamefont {N.}~\bibnamefont
  {Read}}\ and\ \bibinfo {author} {\bibfnamefont {S.}~\bibnamefont {Sachdev}},\
  }\href {\doibase 10.1103/PhysRevLett.66.1773} {\bibfield  {journal} {\bibinfo
   {journal} {Phys. Rev. Lett.}\ }\textbf {\bibinfo {volume} {66}},\ \bibinfo
  {pages} {1773} (\bibinfo {year} {1991})}\BibitemShut {NoStop}%
\bibitem [{\citenamefont {Yang}\ \emph {et~al.}(1993)\citenamefont {Yang},
  \citenamefont {Warman},\ and\ \citenamefont
  {Girvin}}]{Yang-PhysRevLett.70.2641}%
  \BibitemOpen
  \bibfield  {author} {\bibinfo {author} {\bibfnamefont {K.}~\bibnamefont
  {Yang}}, \bibinfo {author} {\bibfnamefont {L.~K.}\ \bibnamefont {Warman}}, \
  and\ \bibinfo {author} {\bibfnamefont {S.~M.}\ \bibnamefont {Girvin}},\
  }\href {\doibase 10.1103/PhysRevLett.70.2641} {\bibfield  {journal} {\bibinfo
   {journal} {Phys. Rev. Lett.}\ }\textbf {\bibinfo {volume} {70}},\ \bibinfo
  {pages} {2641} (\bibinfo {year} {1993})}\BibitemShut {NoStop}%
\bibitem [{\citenamefont {Hastings}(2000)}]{Hastings-PhysRevB.63.014413}%
  \BibitemOpen
  \bibfield  {author} {\bibinfo {author} {\bibfnamefont {M.~B.}\ \bibnamefont
  {Hastings}},\ }\href {\doibase 10.1103/PhysRevB.63.014413} {\bibfield
  {journal} {\bibinfo  {journal} {Phys. Rev. B}\ }\textbf {\bibinfo {volume}
  {63}},\ \bibinfo {pages} {014413} (\bibinfo {year} {2000})}\BibitemShut
  {NoStop}%
\bibitem [{\citenamefont {Elser}(1989)}]{Elser-PhysRevLett.62.2405}%
  \BibitemOpen
  \bibfield  {author} {\bibinfo {author} {\bibfnamefont {V.}~\bibnamefont
  {Elser}},\ }\href {\doibase 10.1103/PhysRevLett.62.2405} {\bibfield
  {journal} {\bibinfo  {journal} {Phys. Rev. Lett.}\ }\textbf {\bibinfo
  {volume} {62}},\ \bibinfo {pages} {2405} (\bibinfo {year}
  {1989})}\BibitemShut {NoStop}%
\bibitem [{\citenamefont {Marston}\ and\ \citenamefont
  {Zeng}(1991)}]{Marston-JAppPhys1991}%
  \BibitemOpen
  \bibfield  {author} {\bibinfo {author} {\bibfnamefont {J.~B.}\ \bibnamefont
  {Marston}}\ and\ \bibinfo {author} {\bibfnamefont {C.}~\bibnamefont {Zeng}},\
  }\href {\doibase http://dx.doi.org/10.1063/1.347830"} {\bibfield  {journal}
  {\bibinfo  {journal} {Journal of Applied Physics}\ }\textbf {\bibinfo
  {volume} {69}},\ \bibinfo {pages} {5962} (\bibinfo {year}
  {1991})}\BibitemShut {NoStop}%
\bibitem [{\citenamefont {Nikolic}\ and\ \citenamefont
  {Senthil}(2003)}]{nikolic-PhysRevB.68.214415}%
  \BibitemOpen
  \bibfield  {author} {\bibinfo {author} {\bibfnamefont {P.}~\bibnamefont
  {Nikolic}}\ and\ \bibinfo {author} {\bibfnamefont {T.}~\bibnamefont
  {Senthil}},\ }\href {\doibase 10.1103/PhysRevB.68.214415} {\bibfield
  {journal} {\bibinfo  {journal} {Phys. Rev. B}\ }\textbf {\bibinfo {volume}
  {68}},\ \bibinfo {pages} {214415} (\bibinfo {year} {2003})}\BibitemShut
  {NoStop}%
\bibitem [{\citenamefont {Singh}\ and\ \citenamefont
  {Huse}(2007)}]{singh-PhysRevB.76.180407}%
  \BibitemOpen
  \bibfield  {author} {\bibinfo {author} {\bibfnamefont {R.~R.~P.}\
  \bibnamefont {Singh}}\ and\ \bibinfo {author} {\bibfnamefont {D.~A.}\
  \bibnamefont {Huse}},\ }\href {\doibase 10.1103/PhysRevB.76.180407}
  {\bibfield  {journal} {\bibinfo  {journal} {Phys. Rev. B}\ }\textbf {\bibinfo
  {volume} {76}},\ \bibinfo {pages} {180407} (\bibinfo {year}
  {2007})}\BibitemShut {NoStop}%
\bibitem [{\citenamefont {Evenbly}\ and\ \citenamefont
  {Vidal}(2010)}]{evenbly-PhysRevLett.104.187203}%
  \BibitemOpen
  \bibfield  {author} {\bibinfo {author} {\bibfnamefont {G.}~\bibnamefont
  {Evenbly}}\ and\ \bibinfo {author} {\bibfnamefont {G.}~\bibnamefont
  {Vidal}},\ }\href {\doibase 10.1103/PhysRevLett.104.187203} {\bibfield
  {journal} {\bibinfo  {journal} {Phys. Rev. Lett.}\ }\textbf {\bibinfo
  {volume} {104}},\ \bibinfo {pages} {187203} (\bibinfo {year}
  {2010})}\BibitemShut {NoStop}%
\bibitem [{\citenamefont {Ran}\ \emph {et~al.}(2007)\citenamefont {Ran},
  \citenamefont {Hermele}, \citenamefont {Lee},\ and\ \citenamefont
  {Wen}}]{ran-PhysRevLett.98.117205}%
  \BibitemOpen
  \bibfield  {author} {\bibinfo {author} {\bibfnamefont {Y.}~\bibnamefont
  {Ran}}, \bibinfo {author} {\bibfnamefont {M.}~\bibnamefont {Hermele}},
  \bibinfo {author} {\bibfnamefont {P.~A.}\ \bibnamefont {Lee}}, \ and\
  \bibinfo {author} {\bibfnamefont {X.-G.}\ \bibnamefont {Wen}},\ }\href
  {\doibase 10.1103/PhysRevLett.98.117205} {\bibfield  {journal} {\bibinfo
  {journal} {Phys. Rev. Lett.}\ }\textbf {\bibinfo {volume} {98}},\ \bibinfo
  {pages} {117205} (\bibinfo {year} {2007})}\BibitemShut {NoStop}%
\bibitem [{\citenamefont {Iqbal}\ \emph {et~al.}(2013)\citenamefont {Iqbal},
  \citenamefont {Becca}, \citenamefont {Sorella},\ and\ \citenamefont
  {Poilblanc}}]{iqbal-PhysRevB.87.060405}%
  \BibitemOpen
  \bibfield  {author} {\bibinfo {author} {\bibfnamefont {Y.}~\bibnamefont
  {Iqbal}}, \bibinfo {author} {\bibfnamefont {F.}~\bibnamefont {Becca}},
  \bibinfo {author} {\bibfnamefont {S.}~\bibnamefont {Sorella}}, \ and\
  \bibinfo {author} {\bibfnamefont {D.}~\bibnamefont {Poilblanc}},\ }\href
  {\doibase 10.1103/PhysRevB.87.060405} {\bibfield  {journal} {\bibinfo
  {journal} {Phys. Rev. B}\ }\textbf {\bibinfo {volume} {87}},\ \bibinfo
  {pages} {060405} (\bibinfo {year} {2013})}\BibitemShut {NoStop}%
\bibitem [{\citenamefont {Depenbrock}\ \emph {et~al.}(2012)\citenamefont
  {Depenbrock}, \citenamefont {McCulloch},\ and\ \citenamefont
  {Schollw\"ock}}]{depenbrock-PhysRevLett.109.067201}%
  \BibitemOpen
  \bibfield  {author} {\bibinfo {author} {\bibfnamefont {S.}~\bibnamefont
  {Depenbrock}}, \bibinfo {author} {\bibfnamefont {I.~P.}\ \bibnamefont
  {McCulloch}}, \ and\ \bibinfo {author} {\bibfnamefont {U.}~\bibnamefont
  {Schollw\"ock}},\ }\href {\doibase 10.1103/PhysRevLett.109.067201} {\bibfield
   {journal} {\bibinfo  {journal} {Phys. Rev. Lett.}\ }\textbf {\bibinfo
  {volume} {109}},\ \bibinfo {pages} {067201} (\bibinfo {year}
  {2012})}\BibitemShut {NoStop}%
\bibitem [{\citenamefont {{Jiang}}\ \emph {et~al.}(2012)\citenamefont
  {{Jiang}}, \citenamefont {{Wang}},\ and\ \citenamefont
  {{Balents}}}]{Jiang-2012NatPh...8..902J}%
  \BibitemOpen
  \bibfield  {author} {\bibinfo {author} {\bibfnamefont {H.-C.}\ \bibnamefont
  {{Jiang}}}, \bibinfo {author} {\bibfnamefont {Z.}~\bibnamefont {{Wang}}}, \
  and\ \bibinfo {author} {\bibfnamefont {L.}~\bibnamefont {{Balents}}},\ }\href
  {\doibase 10.1038/nphys2465} {\bibfield  {journal} {\bibinfo  {journal}
  {Nature Physics}\ }\textbf {\bibinfo {volume} {8}},\ \bibinfo {pages} {902}
  (\bibinfo {year} {2012})},\ \Eprint {http://arxiv.org/abs/1205.4289}
  {arXiv:1205.4289 [cond-mat.str-el]} \BibitemShut {NoStop}%
\bibitem [{\citenamefont {{He}}\ \emph {et~al.}(2016)\citenamefont {{He}},
  \citenamefont {{Zaletel}}, \citenamefont {{Oshikawa}},\ and\ \citenamefont
  {{Pollmann}}}]{2016arXiv161106238H}%
  \BibitemOpen
  \bibfield  {author} {\bibinfo {author} {\bibfnamefont {Y.-C.}\ \bibnamefont
  {{He}}}, \bibinfo {author} {\bibfnamefont {M.~P.}\ \bibnamefont {{Zaletel}}},
  \bibinfo {author} {\bibfnamefont {M.}~\bibnamefont {{Oshikawa}}}, \ and\
  \bibinfo {author} {\bibfnamefont {F.}~\bibnamefont {{Pollmann}}},\
  }\href@noop {} {\bibfield  {journal} {\bibinfo  {journal} {ArXiv e-prints}\ }
  (\bibinfo {year} {2016})},\ \Eprint {http://arxiv.org/abs/1611.06238}
  {arXiv:1611.06238 [cond-mat.str-el]} \BibitemShut {NoStop}%
\bibitem [{\citenamefont {Helton}\ \emph {et~al.}(2007)\citenamefont {Helton},
  \citenamefont {Matan}, \citenamefont {Shores}, \citenamefont {Nytko},
  \citenamefont {Bartlett}, \citenamefont {Yoshida}, \citenamefont {Takano},
  \citenamefont {Suslov}, \citenamefont {Qiu}, \citenamefont {Chung},
  \citenamefont {Nocera},\ and\ \citenamefont {Lee}}]{PhysRevLett.98.107204}%
  \BibitemOpen
  \bibfield  {author} {\bibinfo {author} {\bibfnamefont {J.~S.}\ \bibnamefont
  {Helton}}, \bibinfo {author} {\bibfnamefont {K.}~\bibnamefont {Matan}},
  \bibinfo {author} {\bibfnamefont {M.~P.}\ \bibnamefont {Shores}}, \bibinfo
  {author} {\bibfnamefont {E.~A.}\ \bibnamefont {Nytko}}, \bibinfo {author}
  {\bibfnamefont {B.~M.}\ \bibnamefont {Bartlett}}, \bibinfo {author}
  {\bibfnamefont {Y.}~\bibnamefont {Yoshida}}, \bibinfo {author} {\bibfnamefont
  {Y.}~\bibnamefont {Takano}}, \bibinfo {author} {\bibfnamefont
  {A.}~\bibnamefont {Suslov}}, \bibinfo {author} {\bibfnamefont
  {Y.}~\bibnamefont {Qiu}}, \bibinfo {author} {\bibfnamefont {J.-H.}\
  \bibnamefont {Chung}}, \bibinfo {author} {\bibfnamefont {D.~G.}\ \bibnamefont
  {Nocera}}, \ and\ \bibinfo {author} {\bibfnamefont {Y.~S.}\ \bibnamefont
  {Lee}},\ }\href {\doibase 10.1103/PhysRevLett.98.107204} {\bibfield
  {journal} {\bibinfo  {journal} {Phys. Rev. Lett.}\ }\textbf {\bibinfo
  {volume} {98}},\ \bibinfo {pages} {107204} (\bibinfo {year}
  {2007})}\BibitemShut {NoStop}%
\bibitem [{\citenamefont {Yoshida}\ \emph {et~al.}(2009)\citenamefont
  {Yoshida}, \citenamefont {Okamoto}, \citenamefont {Tayama}, \citenamefont
  {Sakakibara}, \citenamefont {Tokunaga}, \citenamefont {Matsuo}, \citenamefont
  {Narumi}, \citenamefont {Kindo}, \citenamefont {Yoshida}, \citenamefont
  {Takigawa},\ and\ \citenamefont {Hiroi}}]{doi:10.1143/JPSJ.78.043704}%
  \BibitemOpen
  \bibfield  {author} {\bibinfo {author} {\bibfnamefont {H.}~\bibnamefont
  {Yoshida}}, \bibinfo {author} {\bibfnamefont {Y.}~\bibnamefont {Okamoto}},
  \bibinfo {author} {\bibfnamefont {T.}~\bibnamefont {Tayama}}, \bibinfo
  {author} {\bibfnamefont {T.}~\bibnamefont {Sakakibara}}, \bibinfo {author}
  {\bibfnamefont {M.}~\bibnamefont {Tokunaga}}, \bibinfo {author}
  {\bibfnamefont {A.}~\bibnamefont {Matsuo}}, \bibinfo {author} {\bibfnamefont
  {Y.}~\bibnamefont {Narumi}}, \bibinfo {author} {\bibfnamefont
  {K.}~\bibnamefont {Kindo}}, \bibinfo {author} {\bibfnamefont
  {M.}~\bibnamefont {Yoshida}}, \bibinfo {author} {\bibfnamefont
  {M.}~\bibnamefont {Takigawa}}, \ and\ \bibinfo {author} {\bibfnamefont
  {Z.}~\bibnamefont {Hiroi}},\ }\href {\doibase 10.1143/JPSJ.78.043704}
  {\bibfield  {journal} {\bibinfo  {journal} {Journal of the Physical Society
  of Japan}\ }\textbf {\bibinfo {volume} {78}},\ \bibinfo {pages} {043704}
  (\bibinfo {year} {2009})}\BibitemShut {NoStop}%
\bibitem [{\citenamefont {Hiroi}\ \emph {et~al.}(2001)\citenamefont {Hiroi},
  \citenamefont {Hanawa}, \citenamefont {Kobayashi}, \citenamefont {Nohara},
  \citenamefont {Takagi}, \citenamefont {Kato},\ and\ \citenamefont
  {Takigawa}}]{doi:10.1143/JPSJ.70.3377}%
  \BibitemOpen
  \bibfield  {author} {\bibinfo {author} {\bibfnamefont {Z.}~\bibnamefont
  {Hiroi}}, \bibinfo {author} {\bibfnamefont {M.}~\bibnamefont {Hanawa}},
  \bibinfo {author} {\bibfnamefont {N.}~\bibnamefont {Kobayashi}}, \bibinfo
  {author} {\bibfnamefont {M.}~\bibnamefont {Nohara}}, \bibinfo {author}
  {\bibfnamefont {H.}~\bibnamefont {Takagi}}, \bibinfo {author} {\bibfnamefont
  {Y.}~\bibnamefont {Kato}}, \ and\ \bibinfo {author} {\bibfnamefont
  {M.}~\bibnamefont {Takigawa}},\ }\href {\doibase 10.1143/JPSJ.70.3377}
  {\bibfield  {journal} {\bibinfo  {journal} {Journal of the Physical Society
  of Japan}\ }\textbf {\bibinfo {volume} {70}},\ \bibinfo {pages} {3377}
  (\bibinfo {year} {2001})}\BibitemShut {NoStop}%
\bibitem [{\citenamefont {Okamoto}\ \emph {et~al.}(2009)\citenamefont
  {Okamoto}, \citenamefont {Yoshida},\ and\ \citenamefont
  {Hiroi}}]{doi:10.1143/JPSJ.78.033701}%
  \BibitemOpen
  \bibfield  {author} {\bibinfo {author} {\bibfnamefont {Y.}~\bibnamefont
  {Okamoto}}, \bibinfo {author} {\bibfnamefont {H.}~\bibnamefont {Yoshida}}, \
  and\ \bibinfo {author} {\bibfnamefont {Z.}~\bibnamefont {Hiroi}},\ }\href
  {\doibase 10.1143/JPSJ.78.033701} {\bibfield  {journal} {\bibinfo  {journal}
  {Journal of the Physical Society of Japan}\ }\textbf {\bibinfo {volume}
  {78}},\ \bibinfo {pages} {033701} (\bibinfo {year} {2009})}\BibitemShut
  {NoStop}%
\bibitem [{\citenamefont {Aidoudi}\ \emph {et~al.}(2011)\citenamefont
  {Aidoudi}, \citenamefont {Aldous}, \citenamefont {Goff}, \citenamefont {Z.},
  \citenamefont {Attfield}, \citenamefont {Morris},\ and\ \citenamefont
  {Lightfoot}}]{natureChem-Aidoudi2011}%
  \BibitemOpen
  \bibfield  {author} {\bibinfo {author} {\bibfnamefont {F.~H.}\ \bibnamefont
  {Aidoudi}}, \bibinfo {author} {\bibfnamefont {D.~W.}\ \bibnamefont {Aldous}},
  \bibinfo {author} {\bibfnamefont {R.~J.}\ \bibnamefont {Goff}}, \bibinfo
  {author} {\bibfnamefont {S.~M.}\ \bibnamefont {Z.}}, \bibinfo {author}
  {\bibfnamefont {J.~P.}\ \bibnamefont {Attfield}}, \bibinfo {author}
  {\bibfnamefont {R.~E.}\ \bibnamefont {Morris}}, \ and\ \bibinfo {author}
  {\bibfnamefont {P.}~\bibnamefont {Lightfoot}},\ }\href {\doibase
  10.1038/nchem.1129} {\bibfield  {journal} {\bibinfo  {journal} {Nature
  Chemistry}\ }\textbf {\bibinfo {volume} {3}},\ \bibinfo {pages} {801}
  (\bibinfo {year} {2011})}\BibitemShut {NoStop}%
\bibitem [{\citenamefont {Clark}\ \emph {et~al.}(2013)\citenamefont {Clark},
  \citenamefont {Orain}, \citenamefont {Bert}, \citenamefont {De~Vries},
  \citenamefont {Aidoudi}, \citenamefont {Morris}, \citenamefont {Lightfoot},
  \citenamefont {Lord}, \citenamefont {Telling}, \citenamefont {Bonville},
  \citenamefont {Attfield}, \citenamefont {Mendels},\ and\ \citenamefont
  {Harrison}}]{PhysRevLett.110.207208}%
  \BibitemOpen
  \bibfield  {author} {\bibinfo {author} {\bibfnamefont {L.}~\bibnamefont
  {Clark}}, \bibinfo {author} {\bibfnamefont {J.~C.}\ \bibnamefont {Orain}},
  \bibinfo {author} {\bibfnamefont {F.}~\bibnamefont {Bert}}, \bibinfo {author}
  {\bibfnamefont {M.~A.}\ \bibnamefont {De~Vries}}, \bibinfo {author}
  {\bibfnamefont {F.~H.}\ \bibnamefont {Aidoudi}}, \bibinfo {author}
  {\bibfnamefont {R.~E.}\ \bibnamefont {Morris}}, \bibinfo {author}
  {\bibfnamefont {P.}~\bibnamefont {Lightfoot}}, \bibinfo {author}
  {\bibfnamefont {J.~S.}\ \bibnamefont {Lord}}, \bibinfo {author}
  {\bibfnamefont {M.~T.~F.}\ \bibnamefont {Telling}}, \bibinfo {author}
  {\bibfnamefont {P.}~\bibnamefont {Bonville}}, \bibinfo {author}
  {\bibfnamefont {J.~P.}\ \bibnamefont {Attfield}}, \bibinfo {author}
  {\bibfnamefont {P.}~\bibnamefont {Mendels}}, \ and\ \bibinfo {author}
  {\bibfnamefont {A.}~\bibnamefont {Harrison}},\ }\href {\doibase
  10.1103/PhysRevLett.110.207208} {\bibfield  {journal} {\bibinfo  {journal}
  {Phys. Rev. Lett.}\ }\textbf {\bibinfo {volume} {110}},\ \bibinfo {pages}
  {207208} (\bibinfo {year} {2013})}\BibitemShut {NoStop}%
\bibitem [{\citenamefont {Orain}\ \emph {et~al.}(2017)\citenamefont {Orain},
  \citenamefont {Bernu}, \citenamefont {Mendels}, \citenamefont {Clark},
  \citenamefont {Aidoudi}, \citenamefont {Lightfoot}, \citenamefont {Morris},\
  and\ \citenamefont {Bert}}]{PhysRevLett.118.237203}%
  \BibitemOpen
  \bibfield  {author} {\bibinfo {author} {\bibfnamefont {J.-C.}\ \bibnamefont
  {Orain}}, \bibinfo {author} {\bibfnamefont {B.}~\bibnamefont {Bernu}},
  \bibinfo {author} {\bibfnamefont {P.}~\bibnamefont {Mendels}}, \bibinfo
  {author} {\bibfnamefont {L.}~\bibnamefont {Clark}}, \bibinfo {author}
  {\bibfnamefont {F.~H.}\ \bibnamefont {Aidoudi}}, \bibinfo {author}
  {\bibfnamefont {P.}~\bibnamefont {Lightfoot}}, \bibinfo {author}
  {\bibfnamefont {R.~E.}\ \bibnamefont {Morris}}, \ and\ \bibinfo {author}
  {\bibfnamefont {F.}~\bibnamefont {Bert}},\ }\href {\doibase
  10.1103/PhysRevLett.118.237203} {\bibfield  {journal} {\bibinfo  {journal}
  {Phys. Rev. Lett.}\ }\textbf {\bibinfo {volume} {118}},\ \bibinfo {pages}
  {237203} (\bibinfo {year} {2017})}\BibitemShut {NoStop}%
\bibitem [{\citenamefont {He}\ \emph {et~al.}(2014)\citenamefont {He},
  \citenamefont {Sheng},\ and\ \citenamefont {Chen}}]{PhysRevLett.112.137202}%
  \BibitemOpen
  \bibfield  {author} {\bibinfo {author} {\bibfnamefont {Y.-C.}\ \bibnamefont
  {He}}, \bibinfo {author} {\bibfnamefont {D.~N.}\ \bibnamefont {Sheng}}, \
  and\ \bibinfo {author} {\bibfnamefont {Y.}~\bibnamefont {Chen}},\ }\href
  {\doibase 10.1103/PhysRevLett.112.137202} {\bibfield  {journal} {\bibinfo
  {journal} {Phys. Rev. Lett.}\ }\textbf {\bibinfo {volume} {112}},\ \bibinfo
  {pages} {137202} (\bibinfo {year} {2014})}\BibitemShut {NoStop}%
\bibitem [{\citenamefont {Gong}\ \emph {et~al.}()\citenamefont {Gong},
  \citenamefont {Zhu},\ and\ \citenamefont {Sheng}}]{gongCSL-scReports}%
  \BibitemOpen
  \bibfield  {author} {\bibinfo {author} {\bibfnamefont {S.-S.}\ \bibnamefont
  {Gong}}, \bibinfo {author} {\bibfnamefont {W.}~\bibnamefont {Zhu}}, \ and\
  \bibinfo {author} {\bibfnamefont {D.~N.}\ \bibnamefont {Sheng}},\ }\href@noop
  {} {\bibfield  {journal} {\bibinfo  {journal} {Scientific Reports}\ }\textbf
  {\bibinfo {volume} {4}}}\BibitemShut {NoStop}%
\bibitem [{\citenamefont {He}\ and\ \citenamefont
  {Chen}(2015)}]{he-PhysRevLett.114.037201}%
  \BibitemOpen
  \bibfield  {author} {\bibinfo {author} {\bibfnamefont {Y.-C.}\ \bibnamefont
  {He}}\ and\ \bibinfo {author} {\bibfnamefont {Y.}~\bibnamefont {Chen}},\
  }\href {\doibase 10.1103/PhysRevLett.114.037201} {\bibfield  {journal}
  {\bibinfo  {journal} {Phys. Rev. Lett.}\ }\textbf {\bibinfo {volume} {114}},\
  \bibinfo {pages} {037201} (\bibinfo {year} {2015})}\BibitemShut {NoStop}%
\bibitem [{\citenamefont {{L{\"a}uchli}}\ and\ \citenamefont
  {{Moessner}}(2015)}]{lauchli-2015arXiv150404380L}%
  \BibitemOpen
  \bibfield  {author} {\bibinfo {author} {\bibfnamefont {A.~M.}\ \bibnamefont
  {{L{\"a}uchli}}}\ and\ \bibinfo {author} {\bibfnamefont {R.}~\bibnamefont
  {{Moessner}}},\ }\href@noop {} {\bibfield  {journal} {\bibinfo  {journal}
  {ArXiv e-prints}\ } (\bibinfo {year} {2015})},\ \Eprint
  {http://arxiv.org/abs/1504.04380} {arXiv:1504.04380 [cond-mat.quant-gas]}
  \BibitemShut {NoStop}%
\bibitem [{\citenamefont {{Changlani}}\ \emph {et~al.}(2017)\citenamefont
  {{Changlani}}, \citenamefont {{Kochkov}}, \citenamefont {{Kumar}},
  \citenamefont {{Clark}},\ and\ \citenamefont
  {{Fradkin}}}]{2017arXiv170304659C}%
  \BibitemOpen
  \bibfield  {author} {\bibinfo {author} {\bibfnamefont {H.~J.}\ \bibnamefont
  {{Changlani}}}, \bibinfo {author} {\bibfnamefont {D.}~\bibnamefont
  {{Kochkov}}}, \bibinfo {author} {\bibfnamefont {K.}~\bibnamefont {{Kumar}}},
  \bibinfo {author} {\bibfnamefont {B.~K.}\ \bibnamefont {{Clark}}}, \ and\
  \bibinfo {author} {\bibfnamefont {E.}~\bibnamefont {{Fradkin}}},\ }\href@noop
  {} {\bibfield  {journal} {\bibinfo  {journal} {ArXiv e-prints}\ } (\bibinfo
  {year} {2017})},\ \Eprint {http://arxiv.org/abs/1703.04659} {arXiv:1703.04659
  [cond-mat.str-el]} \BibitemShut {NoStop}%
\bibitem [{\citenamefont {{He}}\ \emph {et~al.}(2015)\citenamefont {{He}},
  \citenamefont {{Fuji}},\ and\ \citenamefont
  {{Bhattacharjee}}}]{he-2015arXiv151205381H}%
  \BibitemOpen
  \bibfield  {author} {\bibinfo {author} {\bibfnamefont {Y.-C.}\ \bibnamefont
  {{He}}}, \bibinfo {author} {\bibfnamefont {Y.}~\bibnamefont {{Fuji}}}, \ and\
  \bibinfo {author} {\bibfnamefont {S.}~\bibnamefont {{Bhattacharjee}}},\
  }\href@noop {} {\bibfield  {journal} {\bibinfo  {journal} {ArXiv e-prints}\ }
  (\bibinfo {year} {2015})},\ \Eprint {http://arxiv.org/abs/1512.05381}
  {arXiv:1512.05381 [cond-mat.str-el]} \BibitemShut {NoStop}%
\bibitem [{\citenamefont {Mila}(1998)}]{PhysRevLett.81.2356}%
  \BibitemOpen
  \bibfield  {author} {\bibinfo {author} {\bibfnamefont {F.}~\bibnamefont
  {Mila}},\ }\href {\doibase 10.1103/PhysRevLett.81.2356} {\bibfield  {journal}
  {\bibinfo  {journal} {Phys. Rev. Lett.}\ }\textbf {\bibinfo {volume} {81}},\
  \bibinfo {pages} {2356} (\bibinfo {year} {1998})}\BibitemShut {NoStop}%
\bibitem [{\citenamefont {Zhitomirsky}(2005)}]{PhysRevB.71.214413}%
  \BibitemOpen
  \bibfield  {author} {\bibinfo {author} {\bibfnamefont {M.~E.}\ \bibnamefont
  {Zhitomirsky}},\ }\href {\doibase 10.1103/PhysRevB.71.214413} {\bibfield
  {journal} {\bibinfo  {journal} {Phys. Rev. B}\ }\textbf {\bibinfo {volume}
  {71}},\ \bibinfo {pages} {214413} (\bibinfo {year} {2005})}\BibitemShut
  {NoStop}%
\bibitem [{\citenamefont {Subrahmanyam}(1995)}]{PhysRevB.52.1133}%
  \BibitemOpen
  \bibfield  {author} {\bibinfo {author} {\bibfnamefont {V.}~\bibnamefont
  {Subrahmanyam}},\ }\href {\doibase 10.1103/PhysRevB.52.1133} {\bibfield
  {journal} {\bibinfo  {journal} {Phys. Rev. B}\ }\textbf {\bibinfo {volume}
  {52}},\ \bibinfo {pages} {1133} (\bibinfo {year} {1995})}\BibitemShut
  {NoStop}%
\bibitem [{\citenamefont {Raghu}\ \emph {et~al.}(2000)\citenamefont {Raghu},
  \citenamefont {Rudra}, \citenamefont {Ramasesha},\ and\ \citenamefont
  {Sen}}]{PhysRevB.62.9484}%
  \BibitemOpen
  \bibfield  {author} {\bibinfo {author} {\bibfnamefont {C.}~\bibnamefont
  {Raghu}}, \bibinfo {author} {\bibfnamefont {I.}~\bibnamefont {Rudra}},
  \bibinfo {author} {\bibfnamefont {S.}~\bibnamefont {Ramasesha}}, \ and\
  \bibinfo {author} {\bibfnamefont {D.}~\bibnamefont {Sen}},\ }\href {\doibase
  10.1103/PhysRevB.62.9484} {\bibfield  {journal} {\bibinfo  {journal} {Phys.
  Rev. B}\ }\textbf {\bibinfo {volume} {62}},\ \bibinfo {pages} {9484}
  (\bibinfo {year} {2000})}\BibitemShut {NoStop}%
\bibitem [{\citenamefont {Budnik}\ and\ \citenamefont
  {Auerbach}(2004)}]{PhysRevLett.93.187205}%
  \BibitemOpen
  \bibfield  {author} {\bibinfo {author} {\bibfnamefont {R.}~\bibnamefont
  {Budnik}}\ and\ \bibinfo {author} {\bibfnamefont {A.}~\bibnamefont
  {Auerbach}},\ }\href {\doibase 10.1103/PhysRevLett.93.187205} {\bibfield
  {journal} {\bibinfo  {journal} {Phys. Rev. Lett.}\ }\textbf {\bibinfo
  {volume} {93}},\ \bibinfo {pages} {187205} (\bibinfo {year}
  {2004})}\BibitemShut {NoStop}%
\bibitem [{\citenamefont {Capponi}\ \emph {et~al.}(2004)\citenamefont
  {Capponi}, \citenamefont {L\"auchli},\ and\ \citenamefont
  {Mambrini}}]{PhysRevB.70.104424}%
  \BibitemOpen
  \bibfield  {author} {\bibinfo {author} {\bibfnamefont {S.}~\bibnamefont
  {Capponi}}, \bibinfo {author} {\bibfnamefont {A.}~\bibnamefont {L\"auchli}},
  \ and\ \bibinfo {author} {\bibfnamefont {M.}~\bibnamefont {Mambrini}},\
  }\href {\doibase 10.1103/PhysRevB.70.104424} {\bibfield  {journal} {\bibinfo
  {journal} {Phys. Rev. B}\ }\textbf {\bibinfo {volume} {70}},\ \bibinfo
  {pages} {104424} (\bibinfo {year} {2004})}\BibitemShut {NoStop}%
\bibitem [{\citenamefont {Schaffer}\ \emph {et~al.}(2017)\citenamefont
  {Schaffer}, \citenamefont {Huh}, \citenamefont {Hwang},\ and\ \citenamefont
  {Kim}}]{Schaffer2017}%
  \BibitemOpen
  \bibfield  {author} {\bibinfo {author} {\bibfnamefont {R.}~\bibnamefont
  {Schaffer}}, \bibinfo {author} {\bibfnamefont {Y.}~\bibnamefont {Huh}},
  \bibinfo {author} {\bibfnamefont {K.}~\bibnamefont {Hwang}}, \ and\ \bibinfo
  {author} {\bibfnamefont {Y.~B.}\ \bibnamefont {Kim}},\ }\href {\doibase
  10.1103/PhysRevB.95.054410} {\bibfield  {journal} {\bibinfo  {journal} {Phys.
  Rev. B}\ }\textbf {\bibinfo {volume} {95}},\ \bibinfo {pages} {054410}
  (\bibinfo {year} {2017})}\BibitemShut {NoStop}%
\bibitem [{\citenamefont {Mambrini}\ and\ \citenamefont
  {Mila}(2000)}]{Mambrini2000}%
  \BibitemOpen
  \bibfield  {author} {\bibinfo {author} {\bibfnamefont {M.}~\bibnamefont
  {Mambrini}}\ and\ \bibinfo {author} {\bibfnamefont {F.}~\bibnamefont
  {Mila}},\ }\href {\doibase 10.1007/PL00011071} {\bibfield  {journal}
  {\bibinfo  {journal} {The European Physical Journal B - Condensed Matter and
  Complex Systems}\ }\textbf {\bibinfo {volume} {17}},\ \bibinfo {pages} {651}
  (\bibinfo {year} {2000})}\BibitemShut {NoStop}%
\bibitem [{\citenamefont {Kugel’}\ and\ \citenamefont
  {Khomskii}(1982)}]{Kugel’:1982}%
  \BibitemOpen
  \bibfield  {author} {\bibinfo {author} {\bibfnamefont {K.~I.}\ \bibnamefont
  {Kugel’}}\ and\ \bibinfo {author} {\bibfnamefont {D.~I.}\ \bibnamefont
  {Khomskii}},\ }\href {\doibase 10.1070/PU1982v025n04ABEH004537} {\bibfield
  {journal} {\bibinfo  {journal} {Phys. Usp.}\ }\textbf {\bibinfo {volume}
  {25}},\ \bibinfo {pages} {231} (\bibinfo {year} {1982})}\BibitemShut
  {NoStop}%
\bibitem [{\citenamefont {Penc}\ \emph {et~al.}(2003)\citenamefont {Penc},
  \citenamefont {Mambrini}, \citenamefont {Fazekas},\ and\ \citenamefont
  {Mila}}]{penc2003quantum}%
  \BibitemOpen
  \bibfield  {author} {\bibinfo {author} {\bibfnamefont {K.}~\bibnamefont
  {Penc}}, \bibinfo {author} {\bibfnamefont {M.}~\bibnamefont {Mambrini}},
  \bibinfo {author} {\bibfnamefont {P.}~\bibnamefont {Fazekas}}, \ and\
  \bibinfo {author} {\bibfnamefont {F.}~\bibnamefont {Mila}},\ }\href@noop {}
  {\bibfield  {journal} {\bibinfo  {journal} {Physical Review B}\ }\textbf
  {\bibinfo {volume} {68}},\ \bibinfo {pages} {012408} (\bibinfo {year}
  {2003})}\BibitemShut {NoStop}%
\bibitem [{\citenamefont {Li}\ \emph {et~al.}(1998)\citenamefont {Li},
  \citenamefont {Ma}, \citenamefont {Shi},\ and\ \citenamefont
  {Zhang}}]{PhysRevLett.81.3527}%
  \BibitemOpen
  \bibfield  {author} {\bibinfo {author} {\bibfnamefont {Y.~Q.}\ \bibnamefont
  {Li}}, \bibinfo {author} {\bibfnamefont {M.}~\bibnamefont {Ma}}, \bibinfo
  {author} {\bibfnamefont {D.~N.}\ \bibnamefont {Shi}}, \ and\ \bibinfo
  {author} {\bibfnamefont {F.~C.}\ \bibnamefont {Zhang}},\ }\href {\doibase
  10.1103/PhysRevLett.81.3527} {\bibfield  {journal} {\bibinfo  {journal}
  {Phys. Rev. Lett.}\ }\textbf {\bibinfo {volume} {81}},\ \bibinfo {pages}
  {3527} (\bibinfo {year} {1998})}\BibitemShut {NoStop}%
\bibitem [{\citenamefont {Vernay}\ \emph {et~al.}(2004)\citenamefont {Vernay},
  \citenamefont {Penc}, \citenamefont {Fazekas},\ and\ \citenamefont
  {Mila}}]{PhysRevB.70.014428}%
  \BibitemOpen
  \bibfield  {author} {\bibinfo {author} {\bibfnamefont {F.}~\bibnamefont
  {Vernay}}, \bibinfo {author} {\bibfnamefont {K.}~\bibnamefont {Penc}},
  \bibinfo {author} {\bibfnamefont {P.}~\bibnamefont {Fazekas}}, \ and\
  \bibinfo {author} {\bibfnamefont {F.}~\bibnamefont {Mila}},\ }\href {\doibase
  10.1103/PhysRevB.70.014428} {\bibfield  {journal} {\bibinfo  {journal} {Phys.
  Rev. B}\ }\textbf {\bibinfo {volume} {70}},\ \bibinfo {pages} {014428}
  (\bibinfo {year} {2004})}\BibitemShut {NoStop}%
\bibitem [{\citenamefont {White}(1992)}]{PhysRevLett.69.2863}%
  \BibitemOpen
  \bibfield  {author} {\bibinfo {author} {\bibfnamefont {S.~R.}\ \bibnamefont
  {White}},\ }\href {\doibase 10.1103/PhysRevLett.69.2863} {\bibfield
  {journal} {\bibinfo  {journal} {Phys. Rev. Lett.}\ }\textbf {\bibinfo
  {volume} {69}},\ \bibinfo {pages} {2863} (\bibinfo {year}
  {1992})}\BibitemShut {NoStop}%
\bibitem [{\citenamefont {Schollw\"ock}(2005)}]{RevModPhys.77.259}%
  \BibitemOpen
  \bibfield  {author} {\bibinfo {author} {\bibfnamefont {U.}~\bibnamefont
  {Schollw\"ock}},\ }\href {\doibase 10.1103/RevModPhys.77.259} {\bibfield
  {journal} {\bibinfo  {journal} {Rev. Mod. Phys.}\ }\textbf {\bibinfo {volume}
  {77}},\ \bibinfo {pages} {259} (\bibinfo {year} {2005})}\BibitemShut
  {NoStop}%
\bibitem [{\citenamefont {{Schollw{\"o}ck}}(2011)}]{2011AnPhy.326...96S}%
  \BibitemOpen
  \bibfield  {author} {\bibinfo {author} {\bibfnamefont {U.}~\bibnamefont
  {{Schollw{\"o}ck}}},\ }\href {\doibase 10.1016/j.aop.2010.09.012} {\bibfield
  {journal} {\bibinfo  {journal} {Annals of Physics}\ }\textbf {\bibinfo
  {volume} {326}},\ \bibinfo {pages} {96} (\bibinfo {year} {2011})},\ \Eprint
  {http://arxiv.org/abs/1008.3477} {arXiv:1008.3477 [cond-mat.str-el]}
  \BibitemShut {NoStop}%
\bibitem [{\citenamefont {{Stoudenmire}}\ and\ \citenamefont
  {{White}}(2012)}]{2dDMRGStoudenmireWhite}%
  \BibitemOpen
  \bibfield  {author} {\bibinfo {author} {\bibfnamefont {E.~M.}\ \bibnamefont
  {{Stoudenmire}}}\ and\ \bibinfo {author} {\bibfnamefont {S.~R.}\ \bibnamefont
  {{White}}},\ }\href@noop {} {\bibfield  {journal} {\bibinfo  {journal}
  {Annual Review of Condensed Matter Physics}\ }\textbf {\bibinfo {volume}
  {3}},\ \bibinfo {pages} {111} (\bibinfo {year} {2012})}\BibitemShut {NoStop}%
\bibitem [{\citenamefont {Kj\"all}\ \emph {et~al.}(2013)\citenamefont
  {Kj\"all}, \citenamefont {Zaletel}, \citenamefont {Mong}, \citenamefont
  {Bardarson},\ and\ \citenamefont {Pollmann}}]{PhysRevB.87.235106}%
  \BibitemOpen
  \bibfield  {author} {\bibinfo {author} {\bibfnamefont {J.~A.}\ \bibnamefont
  {Kj\"all}}, \bibinfo {author} {\bibfnamefont {M.~P.}\ \bibnamefont
  {Zaletel}}, \bibinfo {author} {\bibfnamefont {R.~S.~K.}\ \bibnamefont
  {Mong}}, \bibinfo {author} {\bibfnamefont {J.~H.}\ \bibnamefont {Bardarson}},
  \ and\ \bibinfo {author} {\bibfnamefont {F.}~\bibnamefont {Pollmann}},\
  }\href {\doibase 10.1103/PhysRevB.87.235106} {\bibfield  {journal} {\bibinfo
  {journal} {Phys. Rev. B}\ }\textbf {\bibinfo {volume} {87}},\ \bibinfo
  {pages} {235106} (\bibinfo {year} {2013})}\BibitemShut {NoStop}%
\bibitem [{\citenamefont {{L{\"a}uchli}}\ \emph {et~al.}(2016)\citenamefont
  {{L{\"a}uchli}}, \citenamefont {{Sudan}},\ and\ \citenamefont
  {{Moessner}}}]{2016arXiv161106990L}%
  \BibitemOpen
  \bibfield  {author} {\bibinfo {author} {\bibfnamefont {A.~M.}\ \bibnamefont
  {{L{\"a}uchli}}}, \bibinfo {author} {\bibfnamefont {J.}~\bibnamefont
  {{Sudan}}}, \ and\ \bibinfo {author} {\bibfnamefont {R.}~\bibnamefont
  {{Moessner}}},\ }\href@noop {} {\bibfield  {journal} {\bibinfo  {journal}
  {ArXiv e-prints}\ } (\bibinfo {year} {2016})},\ \Eprint
  {http://arxiv.org/abs/1611.06990} {arXiv:1611.06990 [cond-mat.str-el]}
  \BibitemShut {NoStop}%
\bibitem [{\citenamefont {Zauner}\ \emph {et~al.}(2015)\citenamefont {Zauner},
  \citenamefont {Draxler}, \citenamefont {Vanderstraeten}, \citenamefont
  {Degroote}, \citenamefont {Haegeman}, \citenamefont {Rams}, \citenamefont
  {Stojevic}, \citenamefont {Schuch},\ and\ \citenamefont
  {Verstraete}}]{1367-2630-17-5-053002}%
  \BibitemOpen
  \bibfield  {author} {\bibinfo {author} {\bibfnamefont {V.}~\bibnamefont
  {Zauner}}, \bibinfo {author} {\bibfnamefont {D.}~\bibnamefont {Draxler}},
  \bibinfo {author} {\bibfnamefont {L.}~\bibnamefont {Vanderstraeten}},
  \bibinfo {author} {\bibfnamefont {M.}~\bibnamefont {Degroote}}, \bibinfo
  {author} {\bibfnamefont {J.}~\bibnamefont {Haegeman}}, \bibinfo {author}
  {\bibfnamefont {M.~M.}\ \bibnamefont {Rams}}, \bibinfo {author}
  {\bibfnamefont {V.}~\bibnamefont {Stojevic}}, \bibinfo {author}
  {\bibfnamefont {N.}~\bibnamefont {Schuch}}, \ and\ \bibinfo {author}
  {\bibfnamefont {F.}~\bibnamefont {Verstraete}},\ }\href@noop {} {\bibfield
  {journal} {\bibinfo  {journal} {New Journal of Physics}\ }\textbf {\bibinfo
  {volume} {17}},\ \bibinfo {pages} {053002} (\bibinfo {year}
  {2015})}\BibitemShut {NoStop}%
\bibitem [{\citenamefont {{Lieb}}\ \emph {et~al.}(1961)\citenamefont {{Lieb}},
  \citenamefont {{Schultz}},\ and\ \citenamefont
  {{Mattis}}}]{1961AnPhy..16..407L}%
  \BibitemOpen
  \bibfield  {author} {\bibinfo {author} {\bibfnamefont {E.}~\bibnamefont
  {{Lieb}}}, \bibinfo {author} {\bibfnamefont {T.}~\bibnamefont {{Schultz}}}, \
  and\ \bibinfo {author} {\bibfnamefont {D.}~\bibnamefont {{Mattis}}},\ }\href
  {\doibase 10.1016/0003-4916(61)90115-4} {\bibfield  {journal} {\bibinfo
  {journal} {Annals of Physics}\ }\textbf {\bibinfo {volume} {16}},\ \bibinfo
  {pages} {407} (\bibinfo {year} {1961})}\BibitemShut {NoStop}%
\bibitem [{\citenamefont {Oshikawa}(2000)}]{PhysRevLett.84.1535}%
  \BibitemOpen
  \bibfield  {author} {\bibinfo {author} {\bibfnamefont {M.}~\bibnamefont
  {Oshikawa}},\ }\href {\doibase 10.1103/PhysRevLett.84.1535} {\bibfield
  {journal} {\bibinfo  {journal} {Phys. Rev. Lett.}\ }\textbf {\bibinfo
  {volume} {84}},\ \bibinfo {pages} {1535} (\bibinfo {year}
  {2000})}\BibitemShut {NoStop}%
\bibitem [{\citenamefont {Misguich}\ \emph {et~al.}(2002)\citenamefont
  {Misguich}, \citenamefont {Lhuillier}, \citenamefont {Mambrini},\ and\
  \citenamefont {Sindzingre}}]{Misguich2002}%
  \BibitemOpen
  \bibfield  {author} {\bibinfo {author} {\bibfnamefont {G.}~\bibnamefont
  {Misguich}}, \bibinfo {author} {\bibfnamefont {C.}~\bibnamefont {Lhuillier}},
  \bibinfo {author} {\bibfnamefont {M.}~\bibnamefont {Mambrini}}, \ and\
  \bibinfo {author} {\bibfnamefont {P.}~\bibnamefont {Sindzingre}},\ }\href
  {\doibase 10.1140/epjb/e20020078} {\bibfield  {journal} {\bibinfo  {journal}
  {The European Physical Journal B - Condensed Matter and Complex Systems}\
  }\textbf {\bibinfo {volume} {26}},\ \bibinfo {pages} {167} (\bibinfo {year}
  {2002})}\BibitemShut {NoStop}%
\bibitem [{\citenamefont {Hastings}(2005)}]{0295-5075-70-6-824}%
  \BibitemOpen
  \bibfield  {author} {\bibinfo {author} {\bibfnamefont {M.~B.}\ \bibnamefont
  {Hastings}},\ }\href@noop {} {\bibfield  {journal} {\bibinfo  {journal} {EPL
  (Europhysics Letters)}\ }\textbf {\bibinfo {volume} {70}},\ \bibinfo {pages}
  {824} (\bibinfo {year} {2005})}\BibitemShut {NoStop}%
\end{thebibliography}%

\vspace{12pt}

\appendix

\section{Stability of the kagome spin liquid: exact diagonalization of small clusters}
\label{app:ED}

\begin{figure}
\begin{center}
\includegraphics[width=0.9\linewidth]{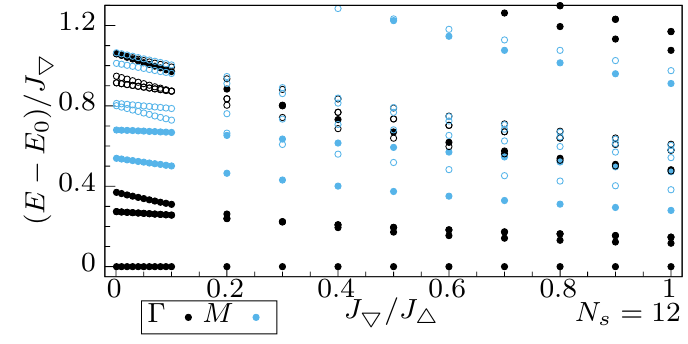}\\
\vspace{6pt}
\includegraphics[width=0.9\linewidth]{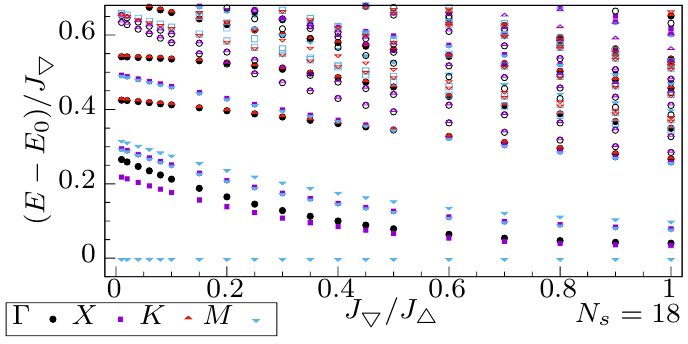}\\
\vspace{6pt}
\includegraphics[width=0.9\linewidth]{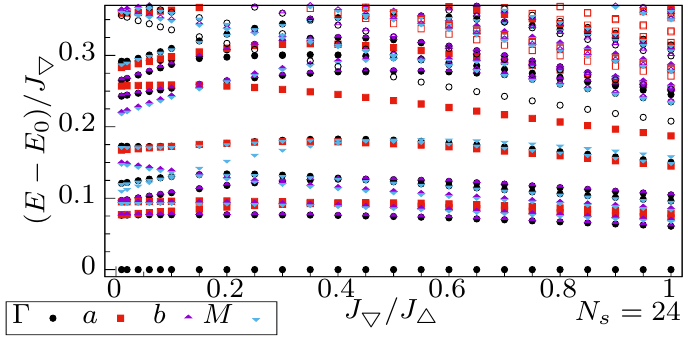}\\
\vspace{6pt}
\includegraphics[width=0.9\linewidth]{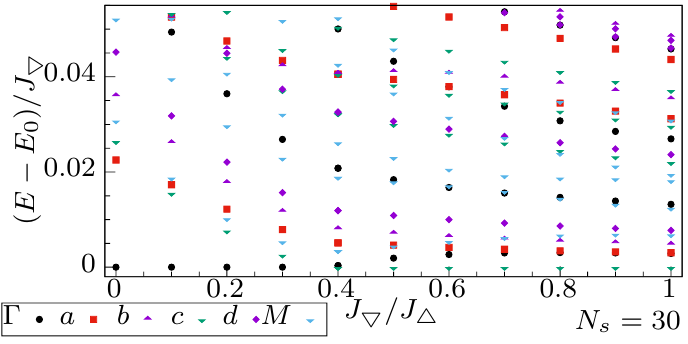}
\caption{Low energy spectrum of several kagome lattice clusters with $N_s$ spins and periodic boundary conditions. The filled (respectively empty) symbols correspond to even (respectively odd) values of $S$. The different colors correspond to different momentum sectors.}
\label{fig:ED app}
\end{center}
\end{figure}

In the main text, we have given the result of the exact diagonalization of the largest toroidal clusters preserving the full $D_3$ symmetry of the Hamiltonian -- i.e. $36$ spins, and $48$ spins in the very anisotropic limit. In Fig.~\ref{fig:ED app}, we give the results obtained for the other clusters: $N_s = 12$ (which preserves $D_3$) and $N_s= 18, 24, 30$ (which break $D_3$). There are significant finite-size effects, but all geometries show a large robustness of the kagome ground state with breathing anisotropy. $N_s =30$ is the smallest size for which we observe a closing of the singlet-singlet gap.

\section{iDMRG results on the $\mathrm{YC2n}-2$ cylinder geometries}
\label{app:DMRG}

\begin{figure}
\includegraphics[width=0.49\linewidth]{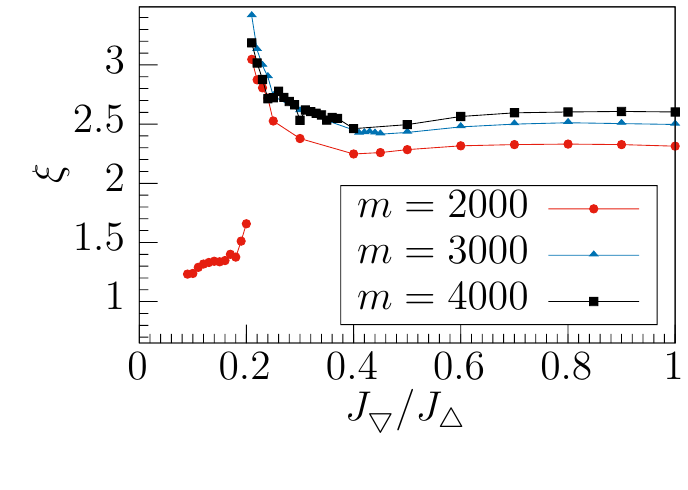}
\includegraphics[width=0.49\linewidth]{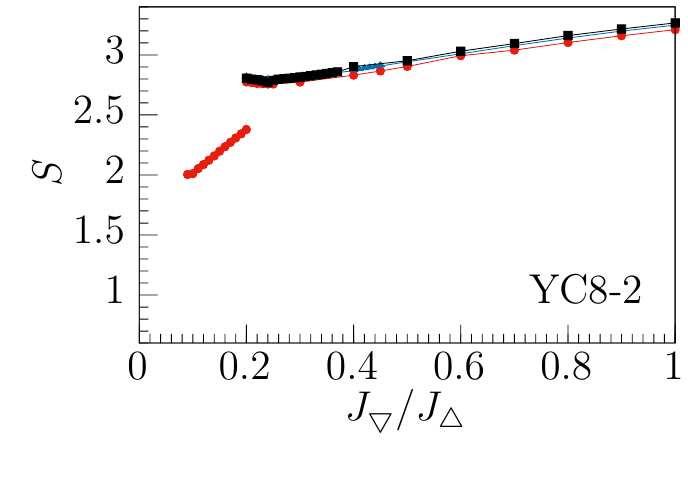}
\includegraphics[width=0.49\linewidth]{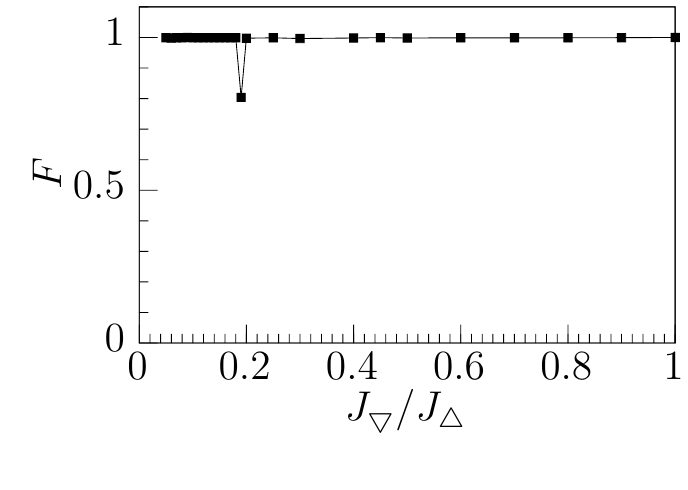}
\includegraphics[width=0.49\linewidth]{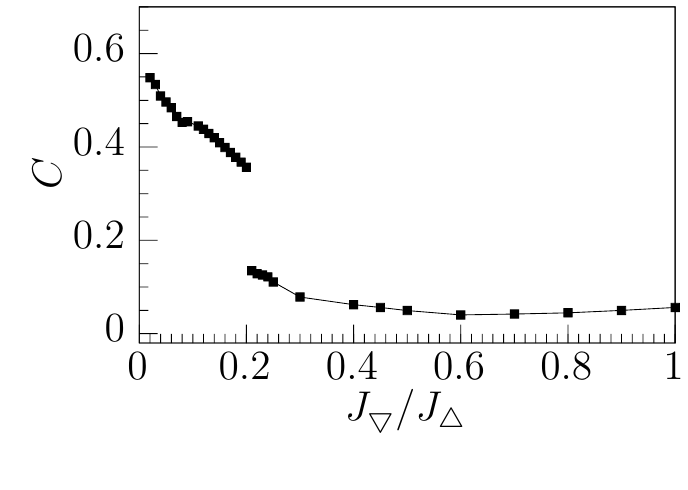}
\caption{Evolution of the correlation length $\xi$, entanglement entropy $S$, and fidelity per ring $F$ and nematic order parameter $C$ Eq.~\eqref{eq:nematic OP} of the ground state of the breathing kagome model Eq.~\eqref{eq:Hamiltonian} on an infinitely long $\mathrm{YC8-2}$ cylinder. Up to $m = 4000$ states were kept in the DMRG simulation.}
\label{fig:YC8-2}
\end{figure}

\begin{figure}
\includegraphics[width=0.49\linewidth]{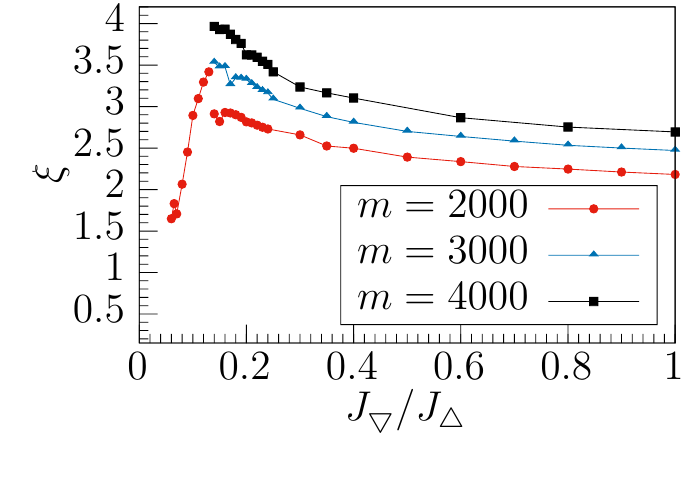}
\includegraphics[width=0.49\linewidth]{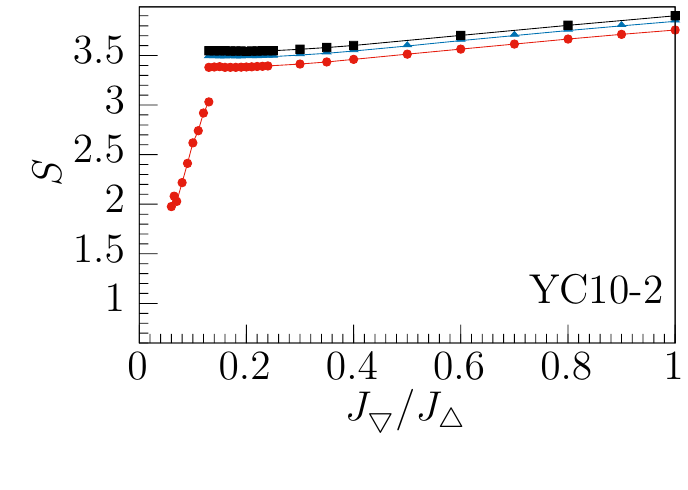}
\includegraphics[width=0.49\linewidth]{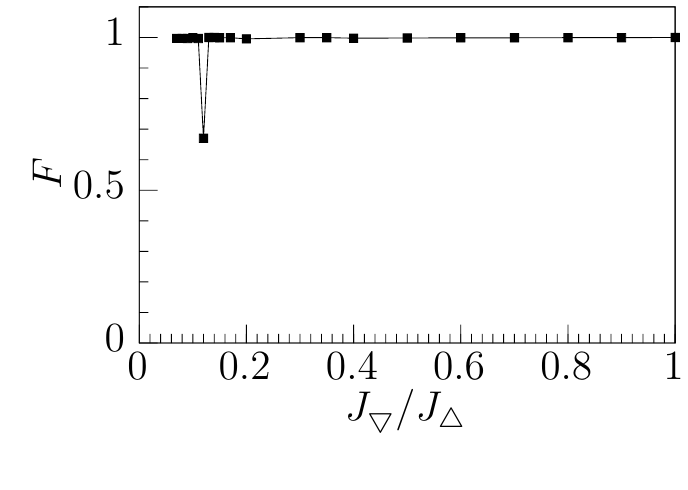}
\includegraphics[width=0.49\linewidth]{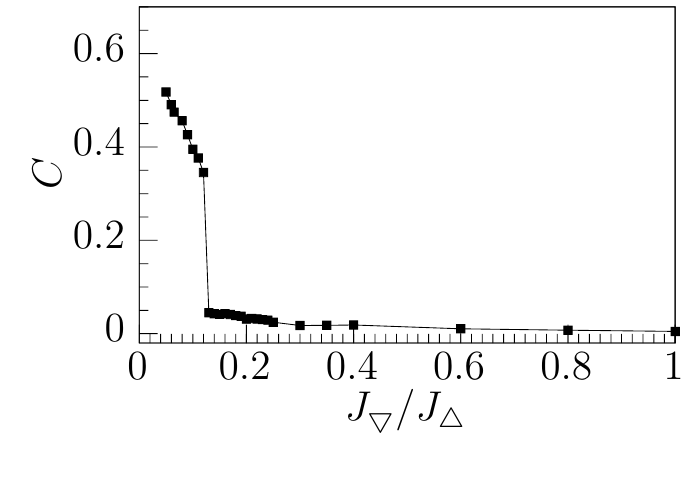}
\caption{Evolution of the correlation length $\xi$, entanglement entropy $S$, and fidelity per ring $F$ and nematic order parameter $C$ Eq.~\eqref{eq:nematic OP} of the ground state of the breathing kagome model Eq.~\eqref{eq:Hamiltonian} on an infinitely long $\mathrm{YC10-2}$ cylinder. Up to $m = 4000$ states were kept in the DMRG simulation.}
\label{fig:YC10-2}
\end{figure}

\begin{figure}
\includegraphics[width=0.49\linewidth]{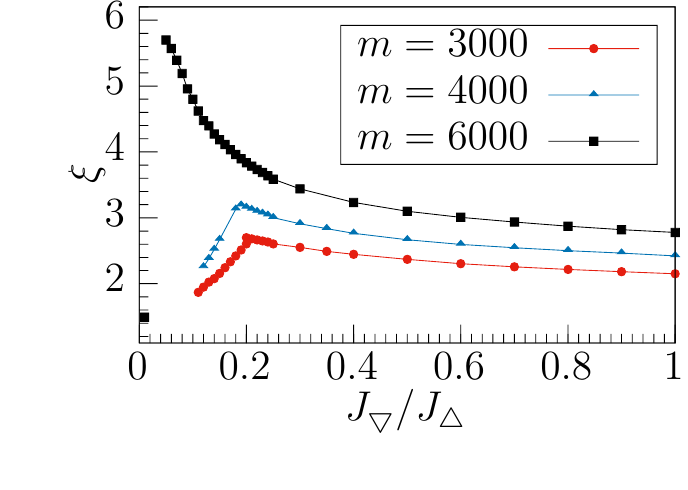}
\includegraphics[width=0.49\linewidth]{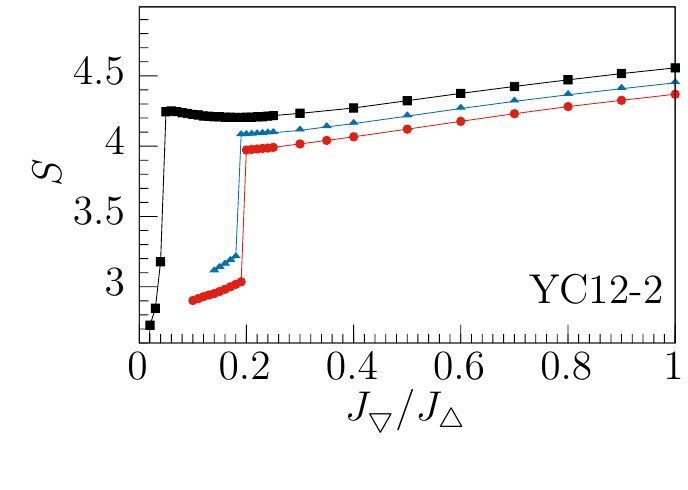}
\includegraphics[width=0.49\linewidth]{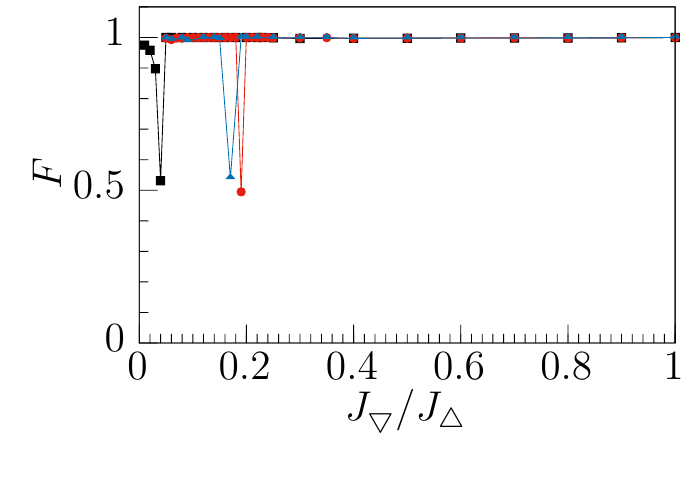}
\includegraphics[width=0.49\linewidth]{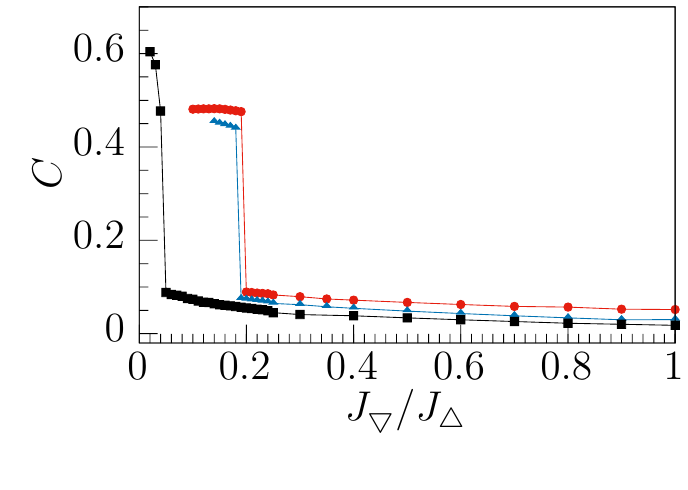}
\caption{Evolution of the correlation length $\xi$, entanglement entropy $S$, and fidelity per ring $F$ and nematic order parameter $C$ Eq.~\eqref{eq:nematic OP} of the ground state of the breathing kagome model Eq.~\eqref{eq:Hamiltonian} on an infinitely long $\mathrm{YC12-2}$ cylinder. Up to $m = 6000$ states were kept in the DMRG simulation.}
\label{fig:YC12-2}
\end{figure}

In this section, we give the result of our iDMRG calculations on the geometries $\mathrm{YC8}-2$ (Fig.~\ref{fig:YC8-2}), $\mathrm{YC10}-2$ (Fig.~\ref{fig:YC10-2}) and $\mathrm{YC12}-2$ (Fig.~\ref{fig:YC12-2}). They complete the results obtained on the $\mathrm{YC8}$ geometry in Fig.~\ref{fig:finite anisotropy}, and confirm the conclusions of Sec.~\ref{sec:stability} regarding the very large stability of the kagome spin liquid with breathing anisotropy. We also track the phase transition to a nematically ordered state.

To evaluate the influence of truncation, we did a scaling of the bond dimension $m$. In the $\mathrm{YC12-2}$ cylinder where the influence of truncation is the largest, we note a shift of the phase transition towards smaller values of $J_{\dwT}^c$ as $m$ becomes larger. The effect of the finite bond dimension is thus to underestimate the stability of the kagome spin liquid.
This phenomenon has a simple explanation: the nematic ground state ($J_{\dwT} < J_{\dwT}^c$) has a much smaller entanglement entropy than its symmetry-preserving ($J_{\dwT} > J_{\dwT}^c$) counterpart, and can be much more efficiently represented in the MPS language. 
 Indeed, the bonds with strong correlations are almost always nearest neighbors in the one-dimensional snake used to map the $2D$ lattice to a $1D$ chain. 
We verified that a different choice for the DMRG snake did not have a significant influence on our results, and especially on the value of $J_{\dwT}^c$. In practice, we explored the values $J_{\dwT} > J_{\upT}$. In the limit of infinite bond dimensions, this is equivalent to $J_{\dwT} < J_{\upT}$, but it can lead to different results for small bond dimensions since it corresponds to a different $1D$-to-$2D$ mapping. For the $\mathrm{YC8}$ geometry, we found the value of the transition to be within a $5\%$ error margin of the value obtained for $J_{\dwT} < J_{\upT}$ after bond dimension extrapolation.

\section{Additional evidence of nematic order in the strongly anisotropic regime: exact diagonalization of toroidal clusters}
\label{app:ED nematic}

\subsection{Quasidegeneracy}
\label{sec:ED GS}

\begin{figure}
\begin{center}
\includegraphics[width=0.9\linewidth]{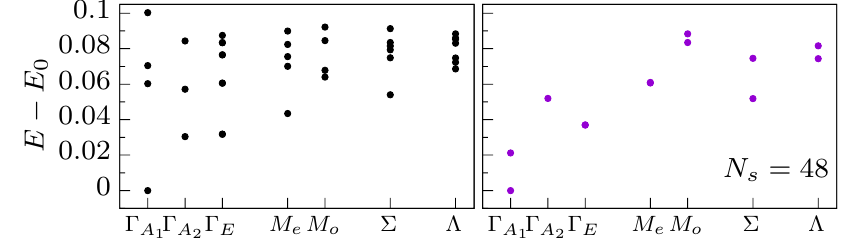}
\end{center}
\caption{Low energy spectrum of a kagome cluster with periodic boundary conditions and $N_s = 48$ spins.
The states are labeled by their momentum sector and by the Mulliken symbol for the irreducible representation of $D_3$ in the $\Gamma$ sector, as well as parity under reflection in the $M$ sector.
We show the low-energy spectrum of the trimerized model $H_\mathrm{eff}$ (left), and, for reference, the low-energy spectrum of the isotropic model (right) as given in Ref.~\onlinecite{2016arXiv161106990L}.}
\label{fig:ed finite anisotropy 48}
\end{figure}

As stated in the main text, in finite size systems, the spontaneous breaking of the $D_3$ symmetry should result in a fourfold quasidegeneracy of the ground state (an $A_1$ state, an $A_2$ state, and two exactly degenerate $E$ states). Focusing on the spectrum of the $D_3$-preserving clusters $N_s = 12, 36, 48$, we note that the three lowest energies in the $\Gamma$ sector correspond indeed respectively to $A_1$, $A_2$ and $E$. These are also the three lowest energies of the system irrespective of momentum sector for $N_s = 48$ (see Fig.~\ref{fig:ed finite anisotropy 48}). This is also the case for $N_s = 12$, although this is perhaps less meaningful given the very small system size. In the three geometries, the finite-size effects are strong, and there is a large splitting between the four low-lying states. 

We focus on the $N_s = 48$ system.
We cannot obtain the full phase diagram (the isotropic point -- which has additional symmetries -- was recently obtained in Ref.~\onlinecite{2016arXiv161106990L} at colossal computational cost). However, we can compare the spectra in the strongly anisotropic limit (obtained by exact diagonalization of $H_\mathrm{eff}$) and at the isotropic point (given in Ref.~\onlinecite{2016arXiv161106990L}). The two corresponding low-energy spectra are given in Fig.~\ref{fig:ed finite anisotropy 48}.
The absolute ground states in either limit fall in the same sector ($\Gamma_{A_1}$), but the first few excitations (which are in the $\Gamma$ sector) have different quantum numbers.
The reorganization of the low energy states within the $\Gamma$ sector is compatible with a phase transition to a nematic phase. While this interpretation cannot be categorically favored, it seems consistent with our observations in other system geometries.

\subsection{Response to a nematic perturbation}

\begin{figure}
\begin{center}
\includegraphics[width = 0.43\linewidth]{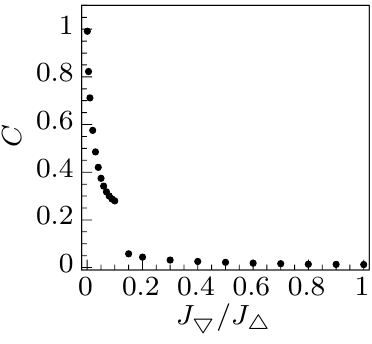}
\includegraphics[width = 0.54\linewidth]{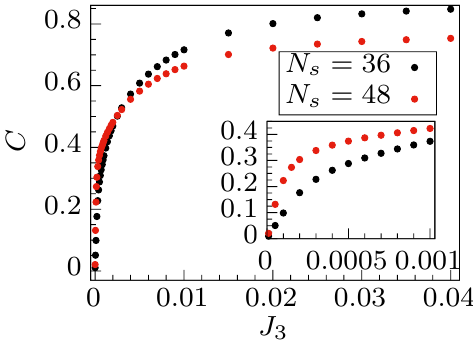}
\caption{Response of the ground state of a finite toroidal cluster (obtained using ED) to a $D_3$-breaking perturbation $H_3$ (see Eq.~\eqref{eq:J3}). The nematic order parameter $C$ is defined in Eq.~\eqref{eq:nematic OP}.
a) The response to the nematic perturbation is measured for various values of $J_{\dwT}/J_{\upT}$ in the $N_s = 36$ system (using the full model Eq.~\eqref{eq:Hamiltonian}), while the perturbation is fixed to $J_3 = 0.001$.
b) $H_3$ is adiabatically added to a system with $N_s$ spins and $J_{\dwT}/J_{\upT} = 0.01$. The ground state of the system was obtained using the effective model eq.~\eqref{eq: triangle hamiltonian}. A zoom-in plot is presented in the inset.}
\label{fig:J3}
\end{center}
\end{figure}

The emergence of a nematic ground state may be hard to detect, since finite-size effects can lift the expected quasidegeneracy to a large extent.
Studying the response of the ground state to a small $D_3$-breaking perturbation term can help us quantify the system's proneness to nematicity. We consider the following perturbation
\begin{equation}
H_3 = J_3 \left(J_{\upT}  \sum_{\langle i,j\rangle  \in \upT'} \mathbf{S_i}\cdot \mathbf{S_j} + J_{\dwT}  \sum_{\langle i,j\rangle  \in \dwT'} \mathbf{S_i}\cdot \mathbf{S_j}\right)
\label{eq:J3}
\end{equation}
where $\upT'$ (respectively $\dwT'$) defines the dimers of the upward (resp. downward) triangle with the largest correlation (full blue and orange bonds in Fig.~\ref{fig:nematic}).
$H_3$ explicitly breaks $D_3$ and favors the formation of a nematic state with a short-range correlation pattern similar to the one observed in the DMRG study of $\mathrm{YC2n}$ cylinders (Fig.~\ref{fig:nematic}).

We present the response of the ground state of the $N_s = 36$ and $N_s = 48$ toroidal clusters to the addition of $H_3$ in Fig.~\ref{fig:J3}. 
We start by evaluating the nematic response~\eqref{eq:nematic OP} across the whole phase diagram. This can be done after diagonalization of the interaction $H + H_3$ in the full Hilbert space for a system of $N_s = 36$ spins. For a small symmetry-breaking term $J_3 = 0.001$, the nematic response is very small at the isotropic point ($C = 0.01$), and remains of the same order of magnitude for $J_{\dwT}/J_{\upT} > 0.1$ ($C<0.06$) as can be seen in Fig.~\ref{fig:J3}a. At $J_{\dwT}/J_{\upT} > 0.1$, we observe a phase transition, with a jump to $C = 0.28$, and an increasingly large response in the strongly anisotropic regime. Note that the position of the jump corresponds to the level crossing in the ground state of the original ($J_3 = 0$) model studied in Sec.~\ref{sec:ED}.
This is in agreement with the much stronger tendency to nematicity in the strongly anisotropic regime already observed on infinite cylinders.

Focusing on the strongly anisotropic regime, we added $H_3$ adiabatically to the original Hamiltonian at $J_{\dwT}/J_{\upT} = 0.01$, using the effective trimerized model for the diagonalization (see Fig.~\ref{fig:J3}b). Using this model allows us to access the next system with $D_3$ symmetry ($N_s = 48$). For both accessible system sizes, there is a finite response even for arbitrarily small perturbations, and a sharp increase of $C$ with increasing $J_3$ (even sharper for the largest system size). This confirms the instability of the ground state to a (seemingly arbitrarily small) nematic perturbation.

\section{Correlations of the ground state of the kagome model with strong breathing anisotropy on toroidal clusters}
\label{app:corr}

\begin{figure*}[h]
\begin{center}
\includegraphics[width = 0.28\linewidth]{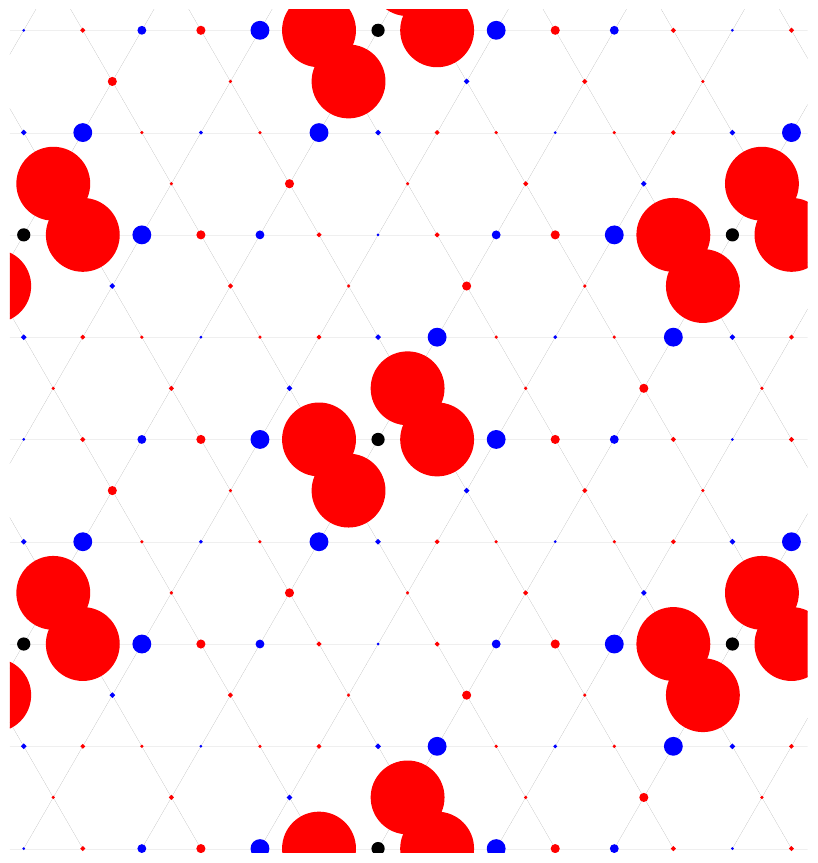}
\includegraphics[width = 0.28\linewidth]{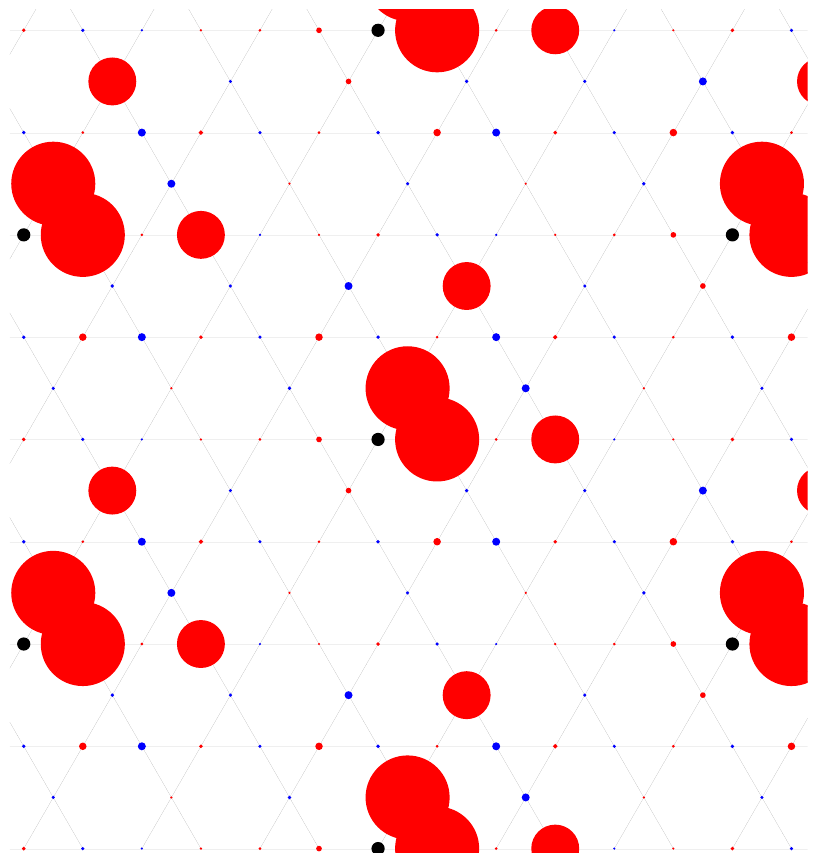}
\includegraphics[width = 0.28\linewidth]{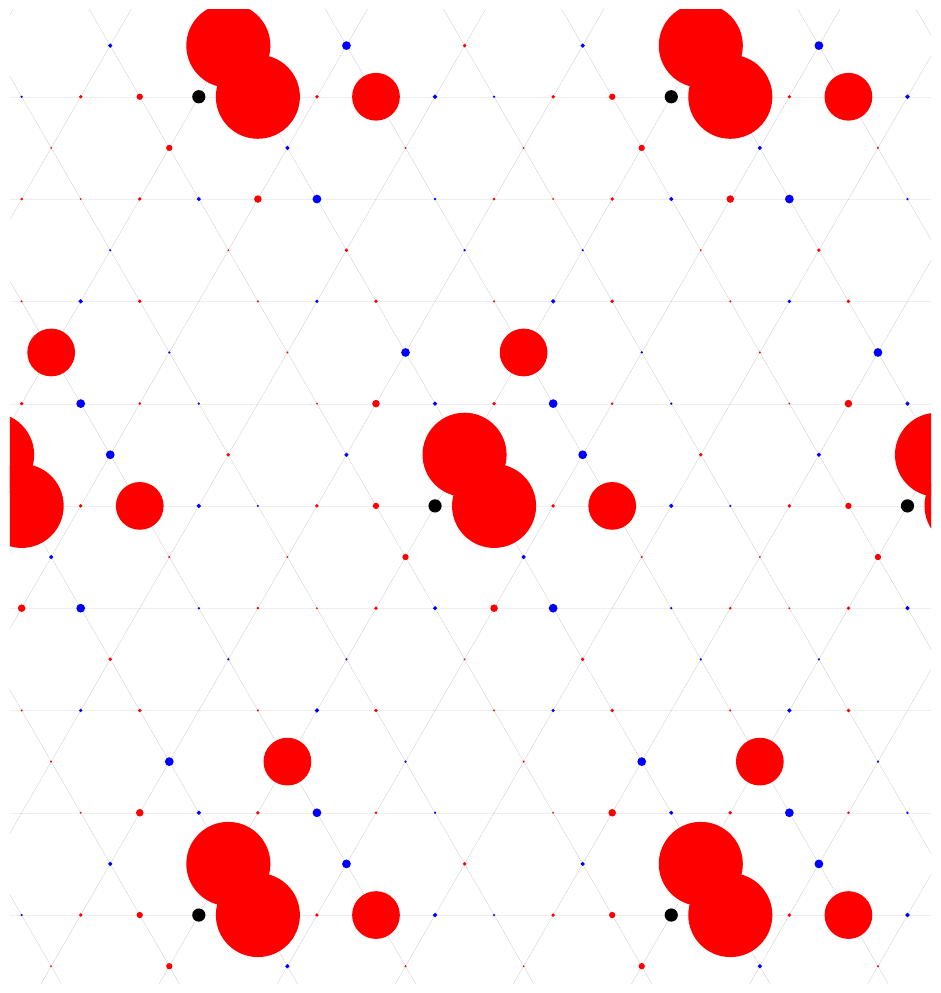}
\caption{Connected spin-spin correlation function $S_c$ of the ground state of the heisenberg model (left) with $N_s = 36$ spins, the trimerized kagome model with $36$ (middle) and $48$ (right) spins. The reference site is indicated in black.}
\label{fig: ED spin-spin}
\end{center}
\end{figure*}

\begin{figure*}[h!]
\begin{center}
\includegraphics[width = 0.28\linewidth]{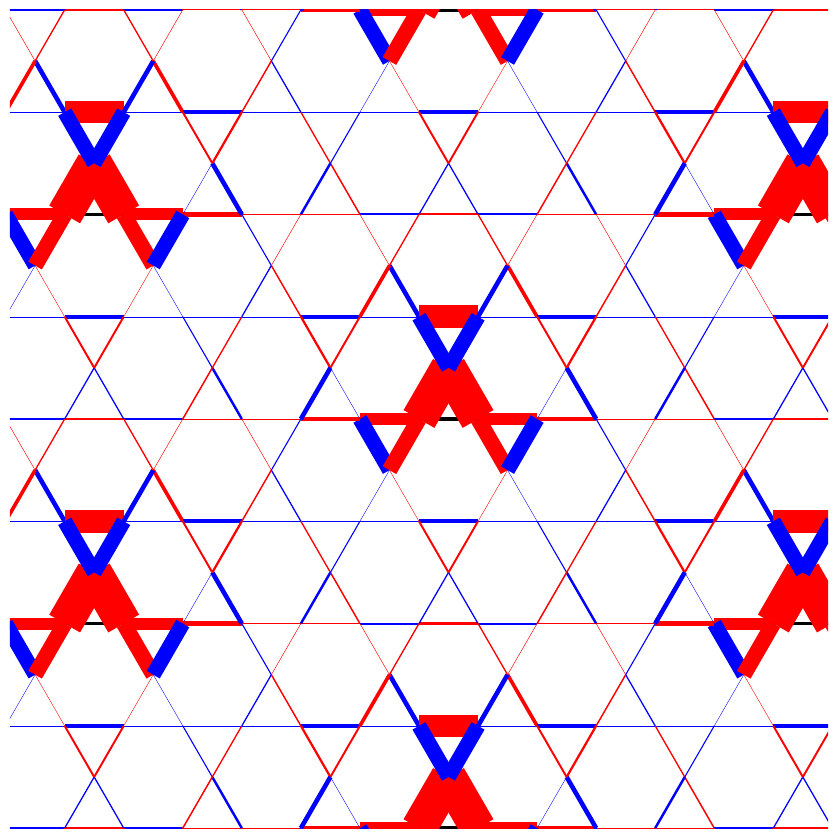}
\includegraphics[width = 0.28\linewidth]{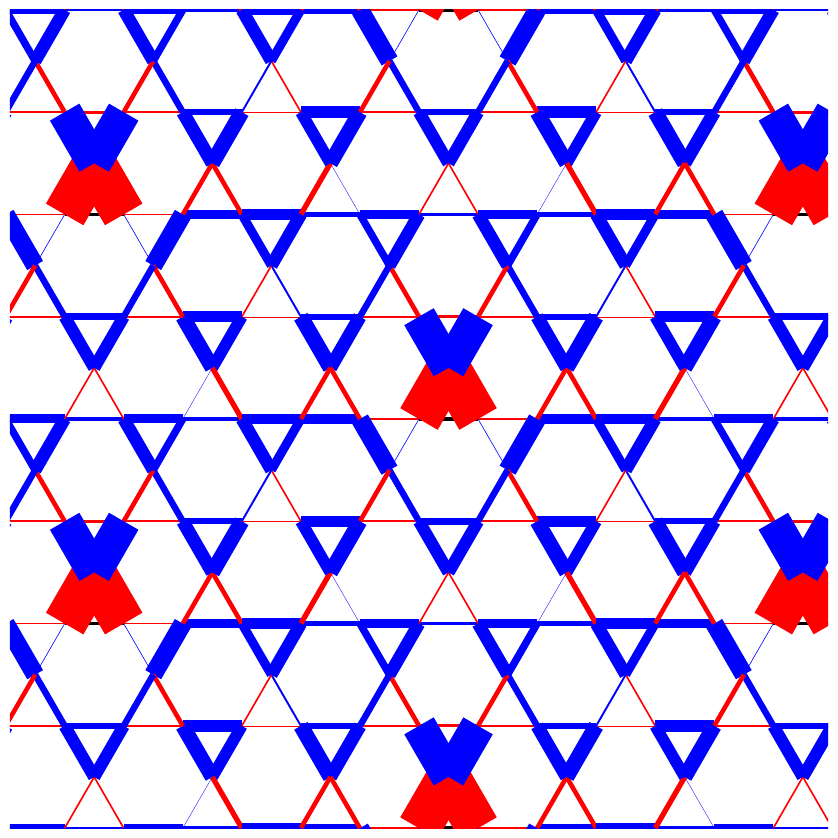}
\includegraphics[width = 0.28\linewidth]{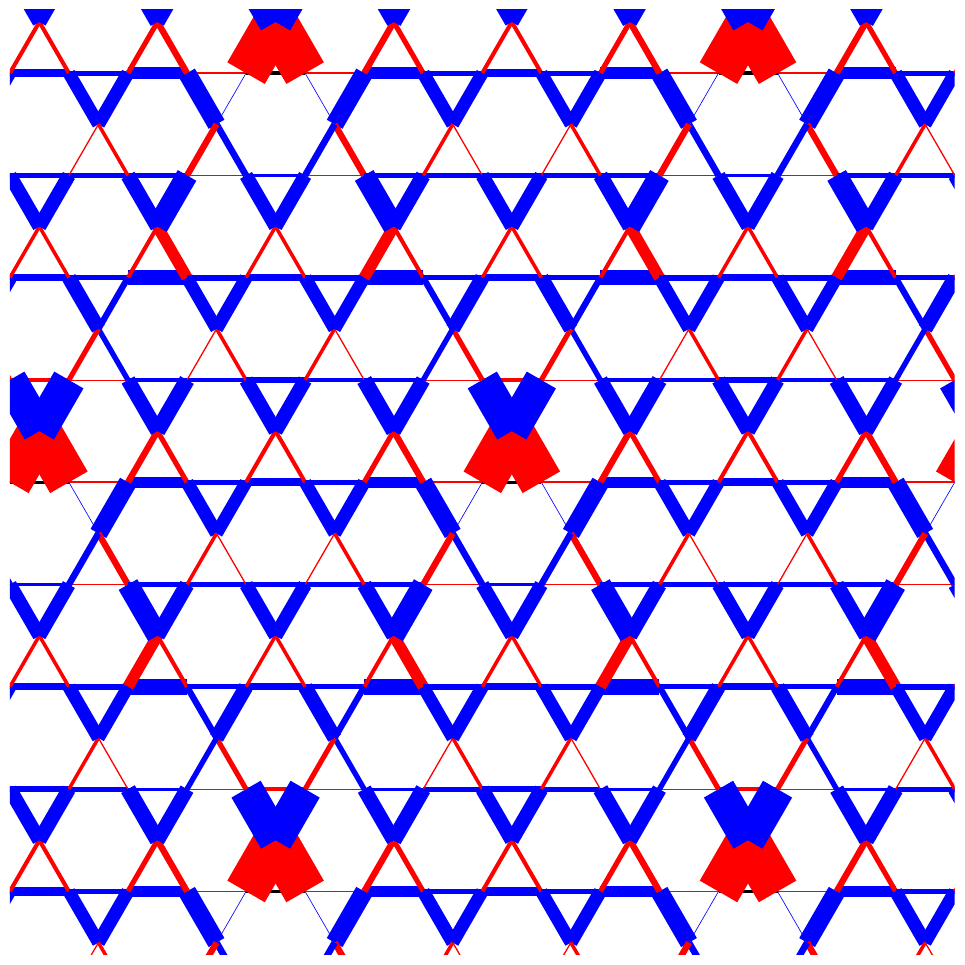}
\caption{Connected dimer-dimer correlation function $D_c$  the ground state of the heisenberg model (left) with $N_s = 36$ spins, the trimerized kagome model with $36$ (middle) and $48$ (right) spins. The reference dimer is indicated in black}
\label{fig: ED dimer}
\end{center}
\end{figure*}

\begin{figure*}[h]
\begin{center}
\includegraphics[width = 0.45\linewidth]{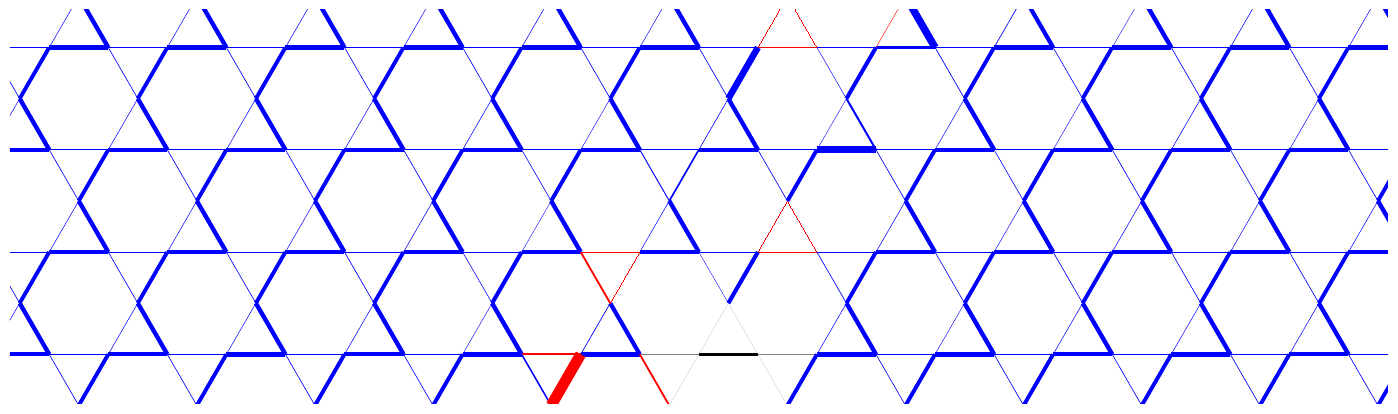}
\hspace{0.5cm}
\includegraphics[width = 0.45\linewidth]{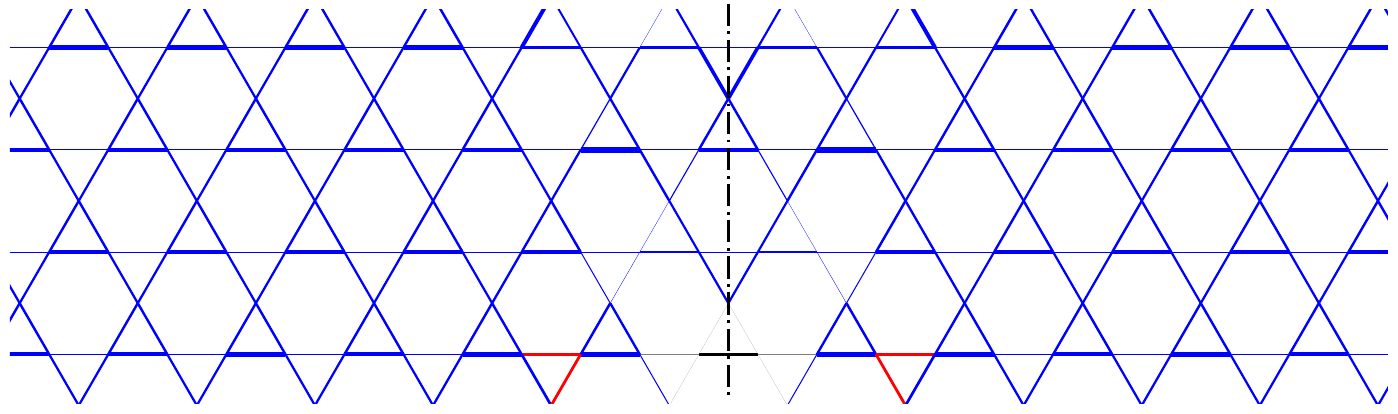}
\caption{Dimer-dimer correlation function $D$ of the ground state of the trimerized kagome model on the $\mathrm{YC8}$ cylinder. The reference dimer is indicated in black. On the right panel, we have symmetrized the correlation function with respect to the black dashed line to facilitate the comparison with the ED results (Fig.~\ref{fig: ED dimer}), where this reflection is an exact symmetry.}
\label{fig: DMRG dimer}
\end{center}
\end{figure*}

In this section, we give the correlation functions of the ground state in the strong breathing anisotropy regime. From the ground state of the effective trimerized model Eq.~\eqref{eq: triangle hamiltonian}, we extracted the spin-spin correlation function
\begin{equation}
S(i) = \langle \mathbf{S_0}\! \cdot\! \mathbf{S_i}\rangle 
\end{equation}
and the connected spin-spin correlation function
\begin{equation}
S_c(i) = \langle \mathbf{S_0}\! \cdot\! \mathbf{S_i}\rangle  - \langle \mathbf{S_0}\rangle \langle \mathbf{S_i}\rangle 
\end{equation}
where $0$ is the index used to label the reference spin.
Similarly, we also extracted the dimer-dimer correlation function and its connected counterpart
\begin{eqnarray}
D(\langle i,j\rangle ) & = & \langle \mathbf{S_0}\! \cdot\! \mathbf{S_1}\ \ \mathbf{S_i}\! \cdot\! \mathbf{S_j}\rangle  \\
D_c(\langle i,j\rangle ) & = & \langle \mathbf{S_0}\! \cdot\! \mathbf{S_1}\ \ \mathbf{S_i}\! \cdot\! \mathbf{S_j}\rangle   - \langle \mathbf{S_0}\! \cdot\! \mathbf{S_1}\rangle  \langle \mathbf{S_i}\! \cdot\! \mathbf{S_j}\rangle  \nonumber
\end{eqnarray}
where $\langle i,j\rangle $ corresponds to two neighboring spins.
The correlation functions are extracted  from the ground state obtained by exact diagonalization of the $N_s = 36, 48$ systems (Figs.~\ref{fig: ED spin-spin} for the spin-spin correlations, \ref{fig: ED dimer} for the dimer-dimer correlations) or iDMRG in the $\mathrm{YC8}$ geometry (Fig.~\ref{fig: DMRG dimer} for the dimer-dimer correlations). In the latter case, the ground state spontaneously breaks the $D_3$ symmetry, and we thus do not use the connected correlation function, which is close to zero at all points. We also give the correlations in the $N_s = 36$ ground state of the isotropic Heisenberg model for reference.

The rapid decay of the spin-spin correlations confirms the disordered nature of the ground state. The dimer-dimer correlations, however, remain finite at relatively large distances in the strong breathing anisotropy regime. Similar patterns can be identified on the torus and infinite cylinder geometries, such as stronger correlations between dimers of the same orientation.

\end{document}